\documentclass[prx,twocolumn,english,superscriptaddress,floatfix,longbibliography]{revtex4-2}

\usepackage{graphicx}
\usepackage{dcolumn}
\usepackage{bm}
\usepackage{amssymb}   
\usepackage{amsmath}
\usepackage{verbatim}
\usepackage{multirow}
\usepackage{xcolor} 
\usepackage{amsmath}
\usepackage{amsfonts}
\usepackage{soul}
\usepackage[pdfencoding=auto,psdextra]{hyperref}
\usepackage{makecell}
\hypersetup{colorlinks=true, pdfstartview=FitV, linkcolor=blue, citecolor=black, plainpages=false, pdfpagelabels=true, urlcolor=blue}
\usepackage[all]{hypcap}

\begin{document}

\title{Modeling phonon-mediated quasiparticle poisoning in superconducting qubit arrays}
\author{E. Yelton}
\affiliation{Department of Physics, Syracuse University, Syracuse, NY 13244-1130, USA} 
\author{C. P. Larson}
\affiliation{Department of Physics, Syracuse University, Syracuse, NY 13244-1130, USA} 
\author{V. Iaia}
\affiliation{Department of Physics, Syracuse University, Syracuse, NY 13244-1130, USA} 
\author{K. Dodge}
\affiliation{Intelligence Community Postdoctoral Research Fellowship Program, Department of Physics, Syracuse University, Syracuse, NY 13244-1130, USA} 
\author{G. La Magna }
\affiliation{JARA Institute for Quantum Information (PGI-11), Forschungszentrum J\"ulich, 52425 J\"ulich, Germany}
\affiliation{Quantum Research Center, Technology Innovation Institute, Abu Dhabi 9639, UAE} 
\author{P. G. Baity}
\affiliation{Computational Science Initiative, Brookhaven National Laboratory, Upton, NY 11973, USA}
\author{I.\,V.~Pechenezhskiy}
\affiliation{Department of Physics, Syracuse University, Syracuse, NY 13244-1130, USA} 
\author{R. McDermott}
\affiliation{Department of Physics, University of Wisconsin-Madison, Madison, Wisconsin 53706, USA}
\author{N.A. Kurinsky}
\affiliation{Kavli Institute for Particle Astrophysics and Cosmology, \\SLAC National Accelerator Laboratory Menlo Park, CA 94025, USA} 
\author{G. Catelani}
\affiliation{JARA Institute for Quantum Information (PGI-11), Forschungszentrum J\"ulich, 52425 J\"ulich, Germany}
\affiliation{Quantum Research Center, Technology Innovation Institute, Abu Dhabi 9639, UAE} 
\author{B. L. T. Plourde}
\affiliation{Department of Physics, Syracuse University, Syracuse, NY 13244-1130, USA} 

\date{\today}

\begin{abstract}
Correlated errors caused by ionizing radiation impacting superconducting qubit chips are problematic for quantum error correction. Such impacts generate quasiparticle (QP) excitations in the qubit electrodes, which temporarily reduce qubit coherence significantly. The many energetic phonons produced by a particle impact travel efficiently throughout the device substrate and generate quasiparticles with high probability, thus causing errors on a large fraction of the qubits in an array simultaneously. We describe a comprehensive strategy for the numerical simulation of the phonon and quasiparticle dynamics in the aftermath of an impact. We compare the simulations with experimental measurements of phonon-mediated QP poisoning and demonstrate that our modeling captures the spatial and temporal footprint of the QP poisoning for various configurations of phonon downconversion structures. We thus present a path forward for the operation of superconducting quantum processors in the presence of ionizing radiation.

\end{abstract}

\maketitle

\section{Introduction}


High-energy particles from background radioactivity or cosmic-ray muons impacting superconducting qubit arrays present a significant challenge for implementing a fault-tolerant quantum processor \cite{McEwen2021,Vepsalainen2020,Wilen2021,Cardani2021,Harrington2024,li2024}. A typical gamma ray from background radioactive contamination hitting the qubit chip deposits energy of order 100~keV in the Si substrate \cite{Wilen2021}. This generates a significant number of electron-hole pairs near the impact site, some of which recombine, while others travel distances of a few hundred $\mu$m before trapping on defects in the Si, thus causing a reconfiguration of the offset-charge environment for any qubits near the impact site \cite{Wilen2021}. Such charge jumps can cause rearrangements of the local two-level system defects near a qubit, 
shifting the qubit frequency and affecting qubit coherence~\cite{Thorbeck2023}. More importantly, the generation of these electron-hole pairs is accompanied by the emission of many energetic athermal phonons, which travel throughout the entire volume of the substrate. Because these phonons generally have energies well above the superconducting energy gap on the device layer, whenever they scatter 
within the superconducting electrodes they will break Cooper pairs and generate nonequilibrium quasiparticles (QPs) with high probability. Transient elevations in QP density in superconducting qubits reduce coherence and enhance the probability of qubit errors. Because of the spread of pair-breaking phonons throughout much of the chip, these QP-induced errors can be correlated across a significant portion of the processor, which current error correction schemes such as the surface code~\cite{Fowler2012} are unable to mitigate.

Software-based approaches to dealing with these correlated errors based on modified error-correcting protocols have been proposed recently \cite{Xu2022,Sane2023,Suzuki2022}. However, due to the added complexity of these schemes, practical approaches for mitigating the QP poisoning in hardware are desirable. While operation in facilities deep underground can shield against cosmic-ray muons \cite{Cardani2021}, and the use of thick Pb layers can protect from environmental radiation sources outside the cryostat \cite{Vepsalainen2020}, these approaches are not always practical; additionally, non-negligible radioactive contamination can also be present in the cryostat and device packaging itself. Some suppression of quasiparticle bursts in superconducting resonators has been achieved by adding small islands of a lower-gap superconductor onto a chip~\cite{Henriques2019,Karatsu2019}, which down-convert the energy from the phonons below the superconducting gap of the device. More recently~\cite{Iaia2022}, some of us demonstrated the use of thick normal-metal islands on the back side of a multi-qubit chip opposite from the device layer, following a scheme proposed in Ref.~\cite{Martinis2021}.
This resulted in a reduction in the rate of two- and three-fold correlated errors by two orders of magnitude.  Understanding the details of the phonon dynamics and QP generation is crucial for further improvements to strategies for mitigating QP poisoning and developing optimal qubit array layouts.

In recent years, a sophisticated numerical tool, GEANT4 Condensed Matter Physics (G4CMP), has been developed for simulating the phonon and charge dynamics in dielectric crystals at millikelvin temperatures, primarily for designing cryogenic dark matter detectors~\cite{Kelsey2023}. Building on the capabilities in GEANT4 for modeling interactions of high-energy particles and matter, the G4CMP package incorporates the relevant solid-state physics to simulate large crystalline substrates, as well as the response of superconducting electrodes on the surfaces for phonon and charge sensing. Here, we apply this software tool to study the phonon and charge dynamics in Si chips containing superconducting qubit arrays. This enables us to explore the effectiveness of different strategies for mitigating phonon-mediated QP poisoning and to quantify the spatial footprint, magnitude, and temporal dynamics of QP generation following an impact event.

\begin{figure}[t!]
\centering
 \includegraphics[width=3.4in]{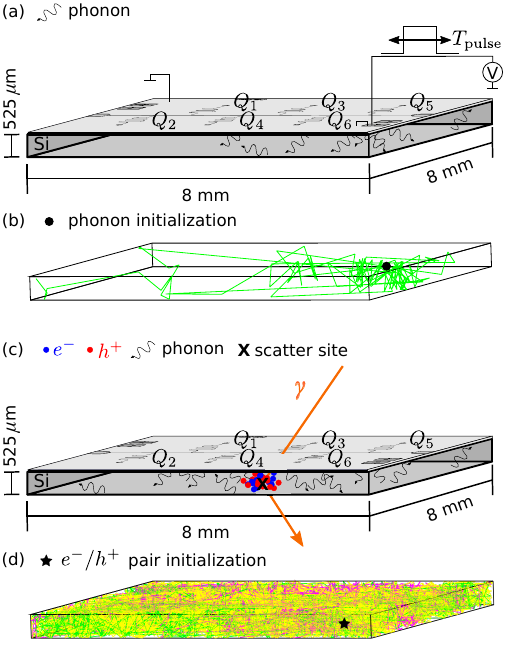}
 \caption{{\bf Experimental device geometry and examples of simulated phonon tracks.} (a) Schematic of physical device geometry under controllable phonon injection via a 
 square bias pulse of length $T_{\rm pulse}$ to an on-chip tunnel junction. (b) G4CMP simulation of a single
 injected phonon in a device with a Nb ground plane and no downconverting film. (c) Diagram of the physical device with the resulting electron-hole pair and phonon production from a $\gamma$-ray impact. (d) Display of 5\% of the simulated phonon tracks resulting from the initialization of a single $e^-/h^+$ pair of energy 3.6~eV. Note that the color for the phonon tracks in (b,d) coincides with the phonon polarization: longitudinal (magenta), slow transverse (yellow), and fast transverse (green).
\label{fig:setup}}
\end{figure}

We describe the typical device layout and the application of G4CMP to simulate phonon dynamics and QP generation in our superconducting qubit system in Sec.~\ref{sec:phonon-modeling}. Here we 
model our experimental measurements of phonon-mediated QP poisoning with direct phonon injection using on-chip tunnel junctions, both from our prior work in Ref.~\cite{Iaia2022} and measurements of new devices with different configurations of phonon mitigation. We use a model for computing the local QP density in the qubit electrodes on the device layer from the numerically modeled phonons to capture the experimental response. In Sec.~\ref{sec:QP-exp} we present measurements of QP charge-parity switching rates, a measure of QP poisoning levels from background events including high-energy particle impacts, for the various devices with different phonon mitigation strategies. Building upon the successful modeling of our direct phonon-injection measurements for various device configurations, in Sec.~\ref{sec:QP-footprint} we simulate the spatial and temporal dynamics of QP generation across a chip following a high-energy particle impact for the various QP mitigation strategies. 
From this modeling, we extract the spatial footprint and temporal dynamics of the QP poisoning for the various device configurations. 



\section{Modeling phonon injection experiments}
\label{sec:phonon-modeling}

G4CMP is a Monte Carlo-based simulation software tool that models electron-hole ($e^-/h^+$) pair production from high-energy particle impacts in dielectric crystals at cryogenic temperatures and their subsequent transport, recombination, and phonon production and the resulting phonon dynamics. Additionally, G4CMP models QP production and scattering in superconducting films from energetic phonons (see Ref.~\cite{Kelsey2023} and Appendix~\hyperref[App:Modeling]{A} for more details). In this work we extend the software to also include phonon scattering in normal metal films (see Appendix~\hyperref[App:Modeling]{A}). QP production is modeled by an energy-flow Monte Carlo simulation between a population of phonons and QPs. This model does not currently account for QP diffusion or recombination, which, as we discuss later, will not significantly limit the relevance of our model.

This work is focused on modeling QP production in four experimental superconducting qubit devices with different levels of phonon mitigation. Each chip consists of an 8~mm~$\times$~8~mm Si substrate of 525~$\mu$m thickness (Fig.~\ref{fig:setup}) with the Si crystal orientation such that the $\langle 110 \rangle$ crystal planes are parallel to the substrate edges. Each device contains an array of up to six charge-sensitive transmon qubits and has a Nb ground plane, resonators, and qubit capacitor islands, plus Al-AlOx-Al Dolan-bridge style Josephson junctions \cite{Dolan1977}. 
\begin{figure}[b!]
\centering
\includegraphics[width=3.4in]{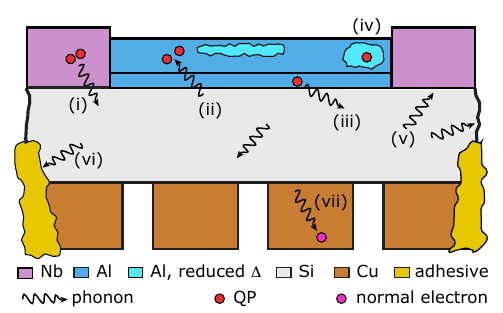}
 \caption{{\bf Schematic of relevant QP and phonon processes.} Features are not to scale. (i) QP recombination and emission of $2\Delta$ phonon. (ii) Pair-breaking and QP generation from phonon with energy $\geq 2\Delta$. (iii) QP relaxation and phonon emission. (iv) QP trapping in region of reduced superconducting gap. (v) Phonon transmission/reflection at interface. (vi) Phonon escape at chip edges. (vii) Phonon scattering and downconversion with conduction electron in normal metal. 
\label{fig:processes}}
\end{figure}
Two of the devices are from Ref.~\cite{Iaia2022}: one with no back-side metallization (non-Cu-A), and one with a 10-$\mu$m-thick back-side Cu film patterned into an array of 200-$\mu$m islands with 50-$\mu$m spacing (10-$\mu$m Cu) for suppressing qubit damping by the lossy transmission-line mode that would be formed otherwise \cite{Martinis2021}. Two new devices have the same layout as those in Ref.~\cite{Iaia2022}: one device also has no back-side metallization (non-Cu-B); the other device has 1-$\mu$m-thick Cu islands with the same geometry as the 10-$\mu$m Cu chip (1-$\mu$m Cu).

Our experimental Josephson elements have Al films of thickness 40~nm (80~nm) for the bottom (top) electrode. Because the superconducting gap of Al depends on the thickness of the film $d$, we use the formula presented in Ref.~\cite{Marchegiani2022}, $\Delta_{\textrm{Al}}(d) = \Delta_{\rm bulk}+ad^{-1}$, where $\Delta_{\rm bulk}$ is 180~$\mu$eV and $a$ is 600~$\mu$eV$\cdot$nm, to estimate $\Delta_{{\rm junc}}$ in the junction electrodes. We take the gap of the simulated Al junction electrode to be the average of the gap in the top and bottom films, 191~$\mu$eV, with a total thickness of 120~nm. 
We simulate the qubit Josephson junction electrodes to be $10\times 10$~$\mu$m$^2$ patches of Al on the top surface of the substrate at the location of each qubit. 
This is larger than the $5\times 1$~$\mu$m$^2$ footprint of our actual qubit junction electrodes on the experimental devices, allowing us to improve the computational efficiency where we gain more 
QP generating events for fewer event trials, assuming the phonon flux is uniform within a small region of the 
device layer. This geometric factor of 20 is accounted for as a scaling of the number of 
QPs generated on each junction throughout our modeling analysis. In our experimental devices, $\sim$99$\%$ of the device layer is covered with the ground plane material, thus we model the top surface as a continuous film of the ground plane 
Nb, with $10\times 10\,\mu\textrm{m}^2$ apertures at the locations of the  Al junction electrode patches. 
In the experimental devices, the size and spacing of the Cu islands results in $\sim$64$\%$ coverage of the back-side surface with metal; our G4CMP simulations match the experimental geometry for these islands.




Our experimental devices are mounted in machined Al sample packages with small amounts of GE varnish adhesive applied at the corners of the substrate. The ground plane of the device layer is connected to the sample packaging with a series of Al wirebonds around the perimeter. Both the GE varnish and wirebonds connect the substrate to the packaging, thus acting as channels for phonons to escape the Si substrate \cite{Martinis2021}. The details of these phonon loss channels are specific to our experimental devices and are not readily modeled in the G4CMP software package. We approximate this phonon loss as a uniform escape probability on the substrate edges (four 525~$\mu$m~$\times$ 8~mm planes). For our simulations, we use the phonon escape probability on the vertical boundaries as a free parameter within the G4CMP simulation to model the recovery timescale following an injection pulse in our experiments. Figure~\ref{fig:processes} depicts the phonon loss at the chip boundaries, as well as the various other phonon and QP processes that are relevant to our modeling effort.

\begin{figure}[t!]
\centering
 \includegraphics[width=3.4in]{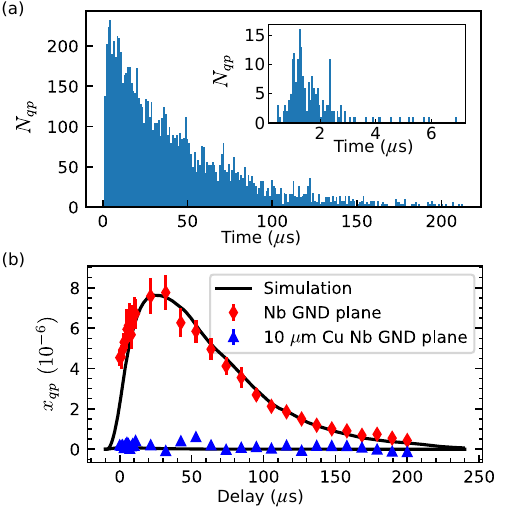}
 \caption{{\bf Simulations of phonon injection measurements.} (a) Distribution of the number of quasiparticles $N_{qp}$ generated on the simulated junction patch at the location of $Q_{4}$ for a device with 
 no back-side metallization (non-Cu-A). Simulated phonon injection includes 10$^{8}$ phonons of energy 2$\Delta_{\rm Al}$. This distribution is used to define the time-dependent QP generation rate $g(t)$ for the QP density modeling. Inset is the number of quasiparticles generated at the $Q_3$ location for a 
 device with 10-$\mu$m-thick islands (10-$\mu$m Cu) from a simulation of $10^9$ phonons. (b) Experimental data points from $\Delta \Gamma_1(t)$ measurements following phonon injection pulse converted to $x_{qp}(t)$ from Ref.~\cite{Iaia2022}; solid lines are resulting simulated curves from G4CMP for $Q_{4}$ on non-Cu-A device (red diamonds) and 10-$\mu$m Cu chip (blue triangles).
\label{fig:phonon-injection}}
\end{figure}

The normalized quasiparticle density $x_{qp}$, defined as the ratio of the volume density of QPs to the volume density of Cooper pairs, can be both 
quantified experimentally and extracted numerically from G4CMP simulations with additional modeling. Here we develop and demonstrate the validity of this modeling. We first consider the phonon injection experiments from Ref.~\cite{Iaia2022}, with a 10-$\mu$s pulse at 1~mV to the on-chip injection junction followed by measurements of qubit $T_1$ at varying times after the end of the pulse. Without structures for phonon downconversion, we observe a substantial increase in the relaxation rate $\Delta\Gamma_1=1/T_1-1/T_1^b$, where $T_1^b$ is the baseline relaxation time with no explicit injection. A quasiparticle tunneling through the qubit Josephson junction can absorb energy from the qubit and cause premature relaxation, therefore, a change in relaxation rate can be directly related to 
the density of excess QPs from the injection: $x_{qp}=\pi\Delta\Gamma_1/\sqrt{2\Delta_{\rm Al}\omega_{01}/\hbar}$, where $\omega_{01}$ is the qubit 0-1 transition frequency \cite{Catelani2011}. In the injection measurements on the non-Cu-A chip, the  $x_{qp}$ level extracted from the measured $\Delta\Gamma_1$ data reaches a maximum of $\sim$7.5~$\times10^{-6}$ approximately $30\,\mu$s after the end of the injection pulse, before recovering back to the baseline level with a time constant of 60~$\mu$s (Fig.~\ref{fig:phonon-injection}).  We model this experiment in the G4CMP simulation by initializing a large number of phonons $N^s_{ph}$ of energy $2\Delta_{\textrm{junc}}$ with a random downward angle from the location of the injector junction [Fig.~\ref{fig:setup}(b)]. From the simulation, we acquire the number of QPs created on each junction electrode $N_{qp}$ versus time $t$, which we use to define a time-dependent QP generation term $g(t)$ in the following model of the normalized QP density $x_{qp}$ in a qubit electrode: 
\begin{equation}
\frac{\textrm{d}x_{qp}}{\textrm{d}t} =-r x_{qp}^{2}-s x_{qp}+g(t),
\label{eq:xqp_model}
\end{equation}
where the QP recombination rate $r$ for Al ranges from $1/\left(100~{\rm ns}\right)$ to $1/\left(10~{\rm ns}\right)$ \cite{Kaplan1976,Ullom1998,Nsanzineza2014,Wang2014} and $s$ is the QP trapping rate due to microscopic variations in the superconducting gap~\cite{Nsanzineza2014,Wang2014}, or possibly the different gaps from the bilayer nature of the Al junction electrodes~\cite{Marchegiani2022,Diamond2022}. 
As we will show, recombination is likely negligible compared to trapping for the relatively low $x_{qp}$ levels in our injection experiments, but we nonetheless include it for completeness, although the precise value of $r$ used in our modeling does not affect the fit significantly. 
As described earlier, G4CMP does not model QP diffusion, and consistently with that, Eq.~(\ref{eq:xqp_model}) is a 0-dimensional treatment of $x_{qp}$ in each electrode, with $g(t)$ determined by $N_{qp}(t)$ following the QP energy cascade in the electrode. We estimate that the QP diffusion time throughout a junction electrode is of the order of 10~ns; since this time scale is much shorter than that associated with trapping (see end of this section), neglecting QP diffusion here is a reasonable approximation (see Appendix~\hyperref[App:Diffusion]{H}).

To relate the QP generation rate $g(t)$ to the phonon injection and subsequent dynamics, we assume the time-dependent generation rate in Eq.~(\ref{eq:xqp_model}) can be written as a convolution of a response function $h(t)$ with the phonon injection rate $I_{ph}(t)$,
\begin{equation}
    g(t) = \int^{t}\!{\rm d}\tau\, h(t-\tau)I_{ph}(\tau).
    \label{eq:g_term_conv}
\end{equation}
The G4CMP simulation initializes all phonons at $t=0$, thus, the simulation results correspond to a delta function phonon injection rate $I_{ph}(t) = N^{s}_{ph}\delta(t)$, where $N^{s}_{ph}$ is the total number of phonons simulated. Figure~\ref{fig:phonon-injection}(a) shows an example histogram of $N_{qp}(t)$ for the $Q_4$ qubit electrode on a device with 
no back-side metallization for $N^{s}_{ph} = 10^8$ phonons, all initialized at $t=0$; the inset to Fig.~\ref{fig:phonon-injection}(a) is a plot of the same quantity for a device with 10-$\mu$m Cu back-side islands, with $N^{s}_{ph} = 10^9$, showing dramatically fewer QPs produced by phonon impacts. The quasiparticle generation rate will be $g(t) = N_{qp}(t)/(n_{cp} V \Delta t )$, where $V$ is the volume of the electrode ($1\times5\times0.12~\mu\textrm{m}^{3}$), $n_{cp}$ ($4\times10^{6}~\mu{\rm m}^{-3}$) is the Cooper pair density, and $\Delta t$ is the bin width of the histogram. Thus, from Eq.~(\ref{eq:g_term_conv}), we obtain the response function as $h(t) = N_{qp}(t)/(n_{cp} V \Delta t N^s_{ph})$. With this result and after discretizing Eq.~(\ref{eq:g_term_conv}), we arrive at the following 
formula relating the generation rate due to an arbitrary injection pulse to the simulated $N_{qp}$:

\begin{equation}
    g(t_i) = \sum_{j<i} \frac{N_{qp}(t_i-t_j)}{n_{cp} V N^{s}_{ph}}I_{ph}(t_j).
    \label{eq:g_term_discrete}
\end{equation}

\begin{figure*}[ht!]
\centering
 \includegraphics[width=6.8in]{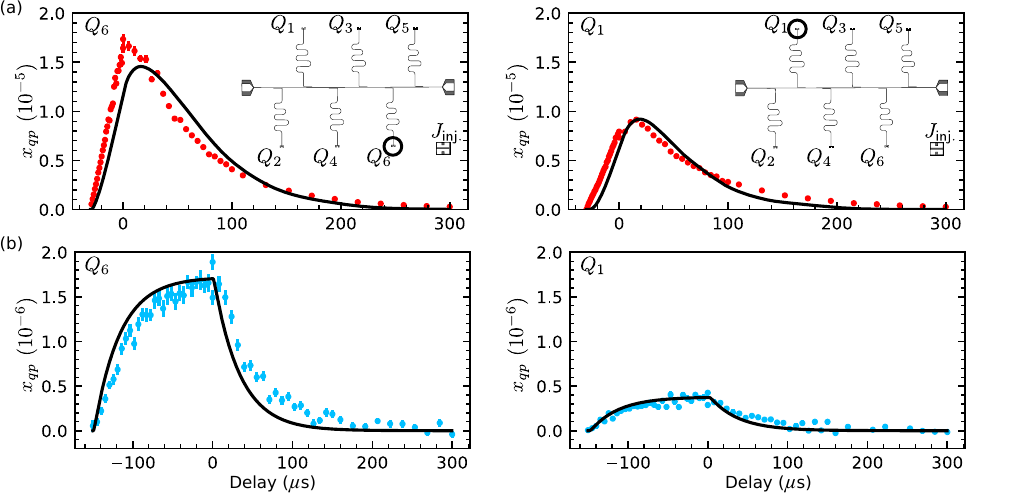}
 \caption{{\bf Experimental and simulated QP response to phonon injection for different qubit locations and phonon mitigation.} $x_{qp}(t)$ from measurements of $\Delta\Gamma_1$ following an injection pulse for qubit closest to injector junction (left) and furthest from injector junction (right) for (a) non-Cu-B chip (red), (b) 1-$\mu$m Cu chip (light blue). Note order-of-magnitude different scales for $x_{qp}$ axis between plots in (a) and (b). Black lines correspond to simulated $x_{qp}(t)$ with fits to experimental data, as described in text.
\label{fig:injection_Nb_1umCu}}
\end{figure*}

To connect our simulation results to experimental measurements and to account for the non-zero width of the injection pulse, we need to define the proper experimental phonon injection rate $I_{ph}(t)$ in Eq.~(\ref{eq:g_term_discrete}). In contrast to the delta function phonon injection rate in the previous paragraph for deriving the response function $h(t)$, the experimental injection pulse has a non-zero duration $T_{\rm pulse}$. For an injection pulse at a junction bias $V_b$, the number of broken Cooper pairs per unit time will be $I_{pair}=V_b/2eR_n$, where $R_n$ is the normal-state resistance of the injector junction. 
The resulting two QPs from each broken Cooper pair have a total energy $eV_b$. In the experiment, $V_b=1$~mV, corresponding to $\sim$5.5$\Delta_{\textrm{Al}}$. 
When $eV_b>4\Delta_{\textrm{Al}}$, at least one of the two QPs from each broken Cooper pair can have sufficient energy to 
emit a $2\Delta_{\textrm{Al}}$ phonon when scattering down to the gap edge. Thus, at $V_b=1$~mV we create pair-breaking phonons via both QP recombination and scattering. To account for this, we simulate an energy cascade of broken pairs of energy 1~meV in a 0-dimensional Monte Carlo scheme, identical to the one described in Appendix~\hyperref[App:Modeling_film]{A.2}. We find that for every broken pair of energy 1~meV, we emit 1.673 pair-breaking phonons ($E_{ph}\geq 2\Delta_{\rm Al}$). With this result and the fact that our injection pulses are square pulses of length $T_{\rm pulse}$, we arrive at the following phonon injection rate, 
\begin{equation}
            I_{ph}(t) = 
            \begin{cases}
                1.673I_{pair}, &  0\leq t \leq T_{\rm pulse} \\
                0, & \textrm{otherwise}.
            \end{cases}
\label{eq:n_inj_pulse}
\end{equation}


We next use Eq.~(\ref{eq:g_term_discrete}) with $I_{ph}(t)$ from Eq.~(\ref{eq:n_inj_pulse}) to compute $g(t)$, which we then include to solve Eq.~(\ref{eq:xqp_model}) 
for $x_{qp}(t)$ using a simple forward Euler method. 
In Fig.~\ref{fig:phonon-injection}(b) we show the simulation results of an injection experiment for one of the qubits on the non-Cu-A chip and a qubit on the 10-$\mu$m Cu device in black compared to the corresponding experimental data from Ref.~\cite{Iaia2022}. For the simulation results, the $t=0$ point is redefined to be the end of the injection pulse of duration $T_{\rm pulse}$ to match the convention in the experiment. In the modeling based on Eq.~(\ref{eq:xqp_model}), we treat the QP trapping rate $s$ as a free parameter for fitting the data. We use a fixed QP recombination rate $r$ of $1/(10~{\rm ns})$.
For the case of $Q_4$ on the non-Cu-A device shown in Fig.~\ref{fig:phonon-injection}(b), we find $s = 4.5(2) \times 10^{-2}~\mu\textrm{s}^{-1}$. From similar analysis for $Q_2$ and $Q_3$, we obtain an average trapping rate of: $s = 4.8(2) \times 10^{-2}~\mu\textrm{s}^{-1}$, which is consistent with results from other QP poisoning measurements of planar multi-layer qubit devices \cite{Leonard2019,Liu2023}. We use a value of $5\times 10^{-2}~\mu\textrm{s}^{-1}$ for $s$ in the modeling of phonon injection for the 10-$\mu$m Cu device. There is a dramatic difference between a 
device without back-side metallization and with 10-$\mu$m Cu islands in response to the phonon injection pulse, consistent with the $N_{qp}(t)$ histogram in the inset of Fig.~\ref{fig:phonon-injection}(a). Thus, the simulation routine we have developed captures this significant variation in device response to the injected phonons.

\begin{figure}[b!]
\centering
 \includegraphics[width=3.4in]{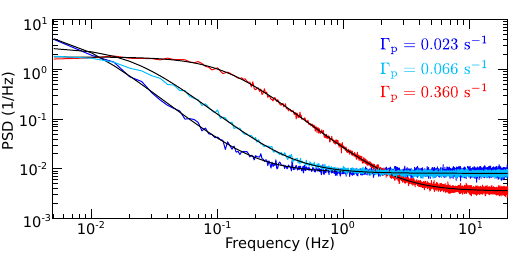}
 \caption{{\bf Experimental QP charge-parity switching power spectral densities.} Measurements with no direct phonon injection for three device configurations: non-Cu-A 
 \cite{Iaia2022} (red), 10~$\mu$m Cu \cite{Iaia2022} (blue), 1-$\mu$m Cu [new] (light blue). Solid black lines are Lorentzian fits to extract $\Gamma_{\textrm{p}}$. 
\label{fig:parity-exp}}
\end{figure}

\begin{figure*}[t!]
\centering
 \includegraphics[width=6.8in]{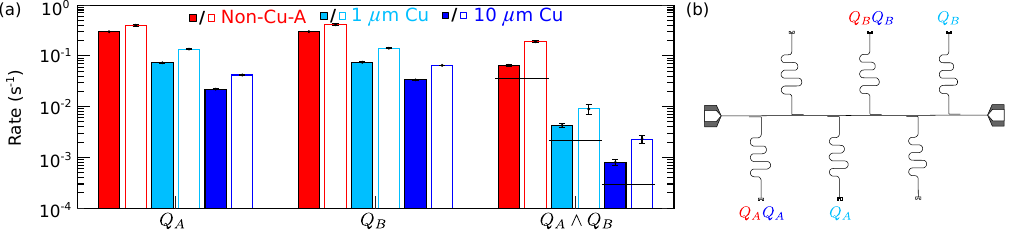}
 \caption{{\bf Experimental QP charge-parity switching rates.} (a) Simultaneous repeated measurements of two qubits --- single-qubit and two-fold QP parity switching rates --- for three 
 different device configurations: non-Cu-A \cite{Iaia2022} (red),  
 10-$\mu$m Cu \cite{Iaia2022} (blue), 1-$\mu$m Cu [new] (light blue). Solid bars depict observed switching rates; hollow bars depict extracted rates for poisoning events (e.g. $\gamma$-ray impacts); horizontal lines indicate the expected background coincidence rates (see Appendix~\hyperref[App:Exp_ChargeParity]{E}). (b) Device layout location of qubit pairs for simultaneous parity switching measurements on each chip.
\label{fig:coincident-parity}}
\end{figure*}

In Fig.~\ref{fig:injection_Nb_1umCu} we present phonon injection data for 
two qubits on the non-Cu-B chip and two on the 1-$\mu$m Cu chip (see Appendix~\hyperref[App:Exp_Inj]{D} for injection data on all 6 qubits on each chip). 
For both devices, we inject phonons from the on-chip tunnel junction indicated in the figure insets and measure $\Delta\Gamma_1$ during and after a phonon injection pulse for the closest ($Q_6$) and furthest ($Q_1$) qubits from the injector junction. Note that the data at negative times on the delay axis correspond to measurements of $\Delta\Gamma_1$ during the injection pulse. In Fig.~\ref{fig:injection_Nb_1umCu}(a) we inject with $T_{\rm pulse}=30\,\mu$s on the non-Cu-B device, and in Fig.~\ref{fig:injection_Nb_1umCu}(b) we inject for $T_{\rm pulse}=150\,\mu$s on the 1-$\mu$m Cu device; the longer pulses for the 1-$\mu$m Cu chip are possible because of the weaker QP poisoning resulting from the phonon mitigation. The longer pulse allows us to observe the saturation of $x_{qp}$ during the pulse, at a level that is roughly an order of magnitude lower than the maximum $x_{qp}$ for injection on the non-Cu-B chip.

The fit curves from the $x_{qp}(t)$ simulations are shown in black (Fig.~\ref{fig:injection_Nb_1umCu}). 
We again observe good agreement between our experimental and simulated $x_{qp}(t)$ response. The average trapping rate for all six qubits on the non-Cu-B device is $5(1)\times10^{-2}~\mu$s$^{-1}$ and for the 1-$\mu$m Cu chip, the average trapping rate is $3.5(8)\times10^{-2}~\mu$s$^{-1}$. The modest variations in trapping rates between the two device configurations 
are likely due to random variations in growth conditions for the junction electrodes during fabrication of the two different wafers. 
The simulations also capture the position dependence in the peak level of $x_{qp}$, including the larger variation in $x_{qp}$ levels between the near and far qubits for the 1-$\mu$m Cu chip compared to the device with no back-side metallization. The lower levels of $x_{qp}$ in the 1-$\mu$m Cu device compared to the non-Cu-B chip under longer injection pulses indicate that a 1-$\mu$m Cu film is effective for down-converting phonons and attenuating their propagation from the injection site. For the qubit closest to the injector junction on the device with 10-$\mu$m thick islands, our modeling predicts a saturation level for a long injection pulse of $x_{qp}\sim10^{-7}$. We were unable to experimentally measure such a small $x_{qp}$ response [Fig.~\ref{fig:phonon-injection}(b)], indicating the 10-$\mu$m film outperforms the 1-$\mu$m one. 

\section{Measurements of QP charge-parity switching}
\label{sec:QP-exp}

As described in Sec.~\ref{sec:phonon-modeling}, we compare three experimental device configurations with different levels of phonon mitigation. 
%
In addition to the measurements of $\Delta\Gamma_1$ under direct phonon injection in the previous section,  another important technique for characterizing QP poisoning involves measuring QP charge-parity switching rates on a qubit island \cite{Riste2013,Christensen2019,Kurter2021,Serniak2018,Pan2022,Connolly2024}. Thus, we perform the same series of measurements as in Ref.~\cite{Iaia2022} of QP charge-parity switching due to background events, including both single-qubit parity switching rates $\Gamma_{\rm p}$ and correlated parity switching for pairs of qubits. 
Experimental details of these measurements are included in Appendix~\hyperref[App:Exp_ChargeParity]{E}. 

In Fig.~\ref{fig:parity-exp} we show the measurements of QP charge-parity switching rates 
$\Gamma_{\rm p}$ from the power spectral density (PSD) of the parity-switching time trace data and the corresponding Lorentzian fits \cite{Riste2013} (Appendix~\hyperref[App:Exp_ChargeParity]{E}). The $\Gamma_{\rm p}$ value for the device with 1-$\mu$m thick Cu back-side islands is $\sim$1/$\left(15~{\rm s}\right)$. This parity switching rate is more than five times lower than the device 
with no back-side metallization (non-Cu-A), but not as low as the device with 10-$\mu$m Cu islands on the back side; $\Gamma_{\rm p}$ for the qubits on the non-Cu-B chip is less than a factor of two different from the non-Cu-A device (Appendix~\hyperref[App:Exp_ChargeParity]{E}). These $\Gamma_{\rm p}$ levels follow the trend we observed for the $x_{qp}$ response on these various devices under controlled phonon injection. 
We note that sources other than background radiation may also contribute to the measured 
$\Gamma_{\rm p}$, 
including IR photons~\cite{Rafferty2021,liu2022,Houzet2019,Diamond2022} and phonon bursts from stresses at the substrate attachment points \cite{Anthony2022}. In our new experiments, the non-Cu-B and 1-$\mu$m Cu devices have identical sample packaging and the same cryogenic environment as the non-Cu-A and 10-$\mu$m Cu chips measured in Ref.~\cite{Iaia2022}. We observe that the rate of large offset charge jumps ($>0.1e$) on the new devices is within a factor of two of the rates reported in Ref.~\cite{Iaia2022}, indicating that these chips all experience a similar radiation background.

\begin{figure*}[!t]
\centering
 \includegraphics[width= 6.8in]{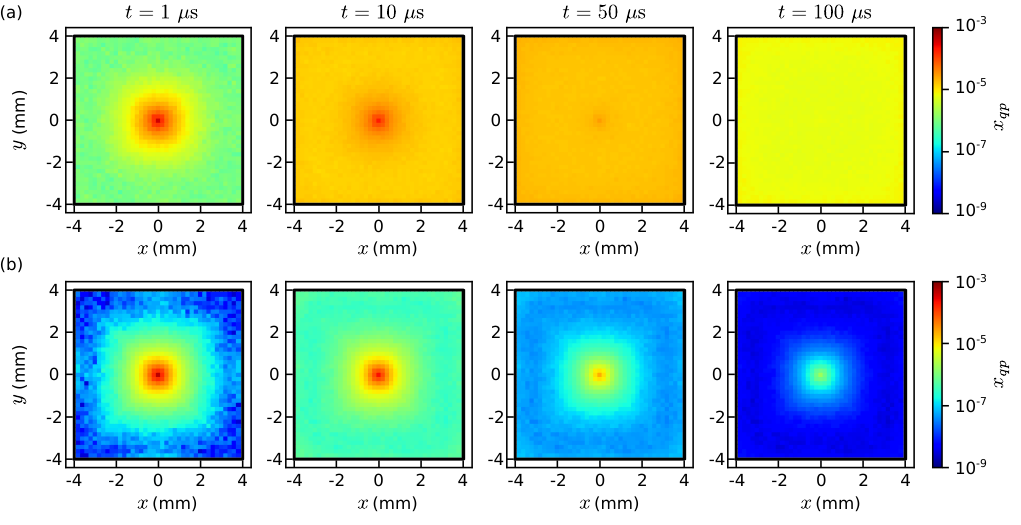}
 \caption{{\bf Visualizing QP poisoning following simulated $\gamma$ impact.} Modeled $x_{qp}(x,y,t)$ across dense qubit array on 200-$\mu$m grid at four different times following initial impact at center of chip. (a) Device with no back-side metallization. (b) Device with 10~$\mu$m-thick Cu islands on back side of chip.
\label{fig:footprint-density-plot}}
\end{figure*}

For each chip, we also monitor simultaneous QP charge-parity switching events for similarly spaced pairs of qubits on each experimental device configuration, which we plot in Fig.~\ref{fig:coincident-parity}. 
We digitize the parity switching data and then count coincident rising/falling edges to extract observed parity switching rates (details in Appendix~\hyperref[App:Exp_ChargeParity]{E}).
The parity switching rates for the single-qubit measurements agree with the rates extracted from the PSD fits in Fig.~\ref{fig:parity-exp}. For two-fold correlated parity switching, the 1-$\mu$m Cu device exhibits a switching rate over 20 times below that of the Nb chip with no explicit phonon downconversion (non-Cu-A).  As demonstrated in Ref.~\cite{Iaia2022}, the 10-$\mu$m Cu device has an even lower two-fold parity switching rate, more than two times lower than the 1-$\mu$m Cu chip. We will show in the following section that these trends are consistent with our modeling of QP poisoning from $\gamma$-ray impacts, including the further improvement in phonon downconversion efficiency upon moving from a 1-$\mu$m thick Cu back-side film to 10~$\mu$m.

\section{QP poisoning footprint in qubit arrays}
\label{sec:QP-footprint}

With the ability to compute $x_{qp}(t)$ in a qubit junction electrode from a simulated injection of pair-breaking phonons, we next 
model $x_{qp}(x,y,t)$ for a hypothetical dense qubit array following a typical $\gamma$-ray impact. A $\gamma$-ray scattering in the device substrate causes an $e^-$ to be directly excited into the conduction band leaving a hole $h^+$ \cite{Kelsey2023}. This energetic $e^-/h^+$ pair ionizes subsequent $e^-/h^+$ pairs in an energy cascade resulting in a population of $N_{eh}^{\gamma}$ $e^-/h^+$ pairs \cite{Ramanathan2020}. We initialize $e^-/h^+$ pairs at a particular location in the device substrate that coincides with an impact site. We simulate a large number of $e^-/h^+$ pairs $N^s_{eh}$ and bin the resulting number of QPs generated on each qubit electrode versus time $N_{qp}(t)$. Similar to our simulations of phonon injection from on-chip tunnel junctions, we use $10\times10~\mu\textrm{m}^2$ patches of Al embedded in openings of the same size in the Nb ground plane to model the qubit junction electrodes. The array of junction electrodes are spaced center-to-center by 200~$\mu$m. We note that this simulated spacing is denser than typical arrays of planar transmons. However, this spacing allows finer resolution of the $x_{qp}$ footprint without altering the ground plane boundary conditions since the grid of $10\times10~\mu\textrm{m}^2$ simulated junctions covers only $\sim0.2\%$ of the area of the top-side of the device substrate. 
We assume all of the $e^-/h^+$ pairs produced by the $\gamma$ impact are initialized at the same time, which we define to be $t=0$. The conversion from the simulated phonon dynamics and QP production to $g(t)$ is more straightforward compared to our earlier modeling of phonon injection with a non-zero junction bias pulse width [Eq.~(\ref{eq:g_term_discrete})], and can instead be expressed as 
\begin{equation}
g(t) =  \frac{N_{qp}(t)}{N^{s}_{eh} n_{cp} V \Delta t}N^{\gamma}_{eh},
\label{eq:gen_term_gamma}
\end{equation}
where 
$N^{\gamma}_{eh}$
is the number of $e^-/h^+$ pairs generated by a single $\gamma$ impact; the remaining parameters are the same as defined in Eq.~(\ref{eq:g_term_discrete}). 
As described in Ref.~\cite{Ramanathan2020}, for recoil energies beyond $\sim$50~eV, each of the $e^-/h^+$ pairs has energy $\sim$3.6~eV, corresponding to a kinetic energy of 2.43~eV after overcoming the Si bandgap. The total number of $e^-/h^+$ pairs generated can be calculated by the ratio of the energy deposited to the energy per pair. We consider a deposition energy of 100~keV, based on the analysis in Ref.~\cite{Wilen2021} that estimated this to be the average characteristic energy deposited from a $\gamma$ impact. 
This leads to $N^{\gamma}_{eh}=\lfloor 100~\textrm{keV}/3.6~\textrm{eV}\rfloor = 27,777$. In order to statistically sample the entire state space of initial charges from the impact energy cascade, we initialize the $N^s_{eh}$ $e^-/h^+$ pairs, each with a randomized momentum direction and with the kinetic energy of the pair randomly distributed between the hole and electron. For a more detailed discussion of this initialization see Appendix~\hyperref[App:Modeling_gamma]{A.1}. 

\begin{figure}[!b]
\centering
 \includegraphics[width= 3.4in]{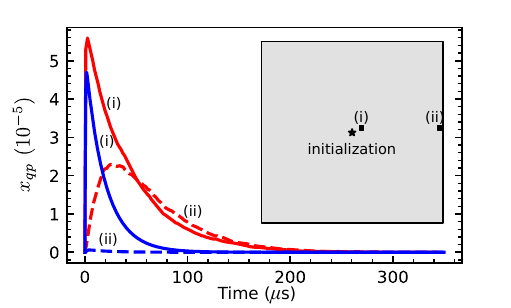}
 \caption{{\bf Simulated $x_{qp}$ response from $\gamma$ impact for two qubit locations.} Plot of $x_{qp}(t)$ response for a location (i) near the impact site (solid lines) and (ii) far from the impact site (dashed lines). 
 Red lines correspond to device with no back-side downconverting film; blue lines correspond to chip with 10-$\mu$m-thick Cu back-side islands.
\label{fig:Edge_center}}
\end{figure}

Following the evolution of the $e^-/h^+$ pairs and the resulting pair-breaking phonons, we solve Eq.~(\ref{eq:xqp_model}) for $x_{qp}(t)$ at each simulated junction electrode location, resulting in $x_{qp}(x,y,t)$ across the top surface of the chip. We run the simulations for various combinations of back-side metallization and metal thickness for the various experimental devices; in addition, we consider a case with superconducting Ti back-side islands. In this last case, we have not yet measured an experimental device, but the small-gap Ti superconductor makes an interesting comparison with the normal metal Cu islands. These simulations are intended to be applicable to a generic superconducting qubit array, however, parameters such as the phonon loss channels and the QP trapping rate $s$ are based on our experimental data for the devices described earlier; these quantities may differ depending on device fabrication and sample packaging details. Figure~\ref{fig:footprint-density-plot}(a,b) shows the resulting $x_{qp}(x,y,t)$ at four instants in time following a $\gamma$ impact in the center of the chip for two simulated devices; one has no back-side metallization and the other has 10-$\mu$m-thick Cu islands on the back side with the same geometry as our 1-$\mu$m Cu and 10-$\mu$m Cu experimental chips. The center of each pixel on the plots coincides with the position of a simulated qubit junction electrode. Immediately following the impact, we observe a significant peak in $x_{qp}$, as high as $6\times10^{-4}$, near the center of the grid for all simulated devices. Without explicit phonon downconversion, such as the chip with no Cu or Ti on the back side, the significantly elevated $x_{qp}$ level spreads rapidly and fills the entire chip, remaining high even 100$\,\mu$s after the impact. In contrast, on the device with 10-$\mu$m Cu back-side metallization, the region of high $x_{qp}$ is confined to a few mm of the impact site, and $x_{qp}$ across the chip recovers to typical background levels in under 100~$\mu$s. The simulated device with 10~$\mu$m of Cu also exhibits features with elevated $x_{qp}$ extending diagonally away from the impact site at short times. This behavior is related to the phonon caustics corresponding to preferred directions of phonon propagation for this particular crystal orientation of the Si substrate (see Appendix~\hyperref[App:Caustic]{F}) \cite{Northrop1979,Martinez2019,Kelsey2023}. 

In Fig.~\ref{fig:Edge_center} we show example $x_{qp}(t)$ responses in the qubit array for both the device with (blue) and without (red) 10-$\mu$m thick Cu islands for a qubit near the impact site $(x=400~\mu{\rm m},y=200~\mu{\rm m})$ (solid line) and a qubit far from the impact site $(x=3800~\mu{\rm m},y=200~\mu{\rm m})$ (dashed line). For the device without a downconverting film we see $x_{qp}$ rise at each qubit location, however, for the device with 10-$\mu$m thick Cu islands we see a similar response for the qubit close to the impact site, although there is a faster recovery, but, importantly, there is almost no rise in $x_{qp}$ for the qubit at the edge of the device. 

\begin{figure}[!t]
\centering
 \includegraphics[width= 3.4in]{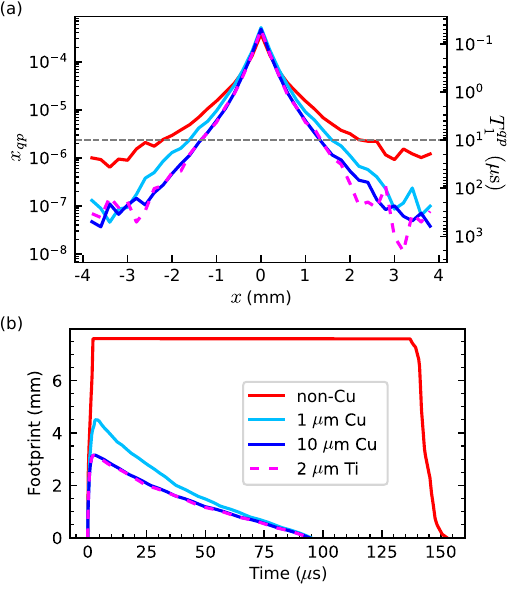}
 \caption{{\bf QP poisoning footprint simulations.} (a) Modeled footprint of $x_{qp}$ and corresponding suppresion of $T_1$ immediately following typical $\gamma$ impact at center of chip on dense qubit arrays for four different combinations of back-side metallization; line-cut taken along $y=0$. (b) Footprint of QP poisoning region extracted from (a) using 10-$\mu$s threshold for $T_1^{qp}$ plotted as a function of time following initial impact.
\label{fig:footprint-linecuts}}
\end{figure}

We compare the profiles of the simulated QP poisoned region immediately after the $\gamma$ impact by plotting horizontal linecuts $x_{qp}(x,y=0,t=1\,\mu{\rm s})$ for the various device configurations [Fig.~\ref{fig:footprint-linecuts}(a)]. While all of the modeled devices exhibit roughly the same peak $x_{qp}$ for qubits close to the impact site, the devices with the most effective phonon downconversion have $x_{qp}$ levels away from the impact that are one order of magnitude lower compared to the chip with no back-side metallization. From the simulated $x_{qp}(x,y,t)$ distributions, we can also compute an instantaneous relaxation time due to QPs at each lattice site; 
we plot the corresponding $T_1^{qp}$ values on the right vertical axis in Fig.~\ref{fig:footprint-linecuts}(a), where we assume $\omega_{01}/2\pi = 5~\textrm{GHz}$. We define the QP poisoning footprint as the spatial extent of the region where $T_1^{qp}$ falls below a threshold, which we choose here to be $10\,\mu$s, indicated by the gray dashed line. In Fig.~\ref{fig:footprint-linecuts}(b) we plot this QP poisoning footprint along the $y=0$ linecut of $T_1^{qp}$ values below this threshold versus time after the impact for the various device configurations. For the chip with no back-side metallization, the footprint rapidly expands to fill the entire chip and only recovers $\sim$ 150~$\mu$s after the impact. Devices with some form of explicit phonon downconversion have much smaller poisoning footprints and recover more quickly. For the most effective phonon mitigation (2-$\mu$m Ti or 10-$\mu$m Cu on back side), the maximum poisoning footprint is  3~mm, only $\sim$14\% of the chip area; this footprint shrinks to zero in under 100~$\mu$s.

The recovery timescale from this modeling is much shorter than the $\sim$25~ms recovery time on the qubit array from Ref.~\cite{McEwen2021}, which had an Al ground plane. However, the recovery timescale we extract from our model is consistent with the measurements in Refs.~\cite{Wilen2021,Thorbeck2023,Iaia2022} that each used a Nb ground plane. Besides the differences in ground plane metal, our modeling indicates at least two key factors that can influence this recovery timescale: 
the QP trapping rate $s$ in the junction electrodes and the phonon loss probability at the chip boundaries (Appendix~\hyperref[App:Exp_Inj]{D} and Fig.~\ref{fig:Trap_Wall_Abs}), which can depend on details of the fabrication and sample packaging. 
In the present work, we have not considered devices with an Al ground plane. QPs have a significantly larger diffusion footprint in an Al film compared to Nb (see Appendix~\hyperref[App:Diffusion]{H}). Thus, one would need to account for QP diffusion in the ground plane in order to extend our modeling to devices with an Al ground plane. 

\section{Conclusion}

We have demonstrated a realistic model of phonon-mediated QP poisoning in superconducting qubit arrays using G4CMP. Our simulations capture the behavior from our experiments of direct phonon injection with on-chip tunnel junctions for several different configurations of back-side metallization. We also observe that the simulated size and recovery of the QP poisoning footprint after a $\gamma$ impact is consistent with our measured trends of multi-qubit coincident charge-parity switching rates due to background radiation, which is likely dominated by $\gamma$-rays, on the different experimental device configurations. Thus, we are able to quantify the effectiveness of various strategies for downconverting the energy from pair-breaking phonons. By extending our modeling approach to a dense grid of hypothetical qubits and simulating the burst of phonons following a $\gamma$ impact, we are able to characterize the footprint of QP poisoning, capturing dramatic differences between the various approaches to phonon mitigation. Our modeling and experimental results thus inform designs for future superconducting qubit arrays and phonon-based sensors of rare events.


\section{Acknowledgments}
\noindent This work is supported by the U.S. Government under ARO grant W911NF-22-1-0257. Fabrication was performed in part at the Cornell NanoScale Facility, a member of the National Nanotechnology Coordinated Infrastructure (NNCI), which is supported by the National Science Foundation (Grant NNCI-2025233). E.Y., C.P.L., P.B., R.M., and B.L.T.P acknowledge partial support by the Laboratory for Physical Sciences through Strategic Partnership Project EAOC0167012.
K.D. acknowledges support by an appointment to the Intelligence Community Postdoctoral Research
Fellowship Program at Syracuse University, administered by Oak Ridge Institute for Science
and Education through an interagency agreement between the U.S. Department of Energy and the Office of
the Director of National Intelligence. \\



\setcounter{section}{0}

\renewcommand{\thesubsection}{\arabic{subsection}}

\section*{Appendix A: G4CMP Modeling}
\label{App:Modeling}
\setcounter{subsection}{0}

As mentioned in the main text, G4CMP models the generation and transport of both $e^-/h^+$ pairs and phonons from ionizing radiation in crystals at cryogenic temperatures. For a detailed overview of the capabilities of this software, we refer the reader to Ref.~\cite{Kelsey2023}. In this appendix, we discuss details relevant to modeling $e^-/h^+$ pairs and the subsequent phonon burst generated from a $\gamma$-ray impact event. In addition, we discuss the film boundary conditions that capture phonon downconversion and QP production.

\subsection{$\gamma$-ray Impacts}
\label{App:Modeling_gamma}

Previous work using GEANT4/G4CMP to model a device with a similar geometry, packaging, and cryostat environment found that a $\gamma$-ray of energy $\sim$ 1~MeV deposits on average $\sim$ 100~keV in the substrate, while cosmic-ray muons and their secondary $\gamma$-rays deposit $\sim$ 460~keV \cite{Wilen2021}. Impacts due to photons or high-energy electrons will directly excite an electron $e^-$ into the conduction band of Si, leaving a hole $h^+$ without depositing energy via lattice vibrations (prompt phonons) \cite{Kelsey2023}. This energetic $e^-/h^+$ pair, the so-called initial hot carrier, ionizes subsequent $e^-/h^+$ pairs in an energy cascade where the total number of $e^-/h^+$ pairs created is $N_{eh} = E_r/\epsilon_{eh}(E_r)$, where $E_r$ is the recoil energy deposited from an impact and $\epsilon_{eh}$ is the average $e^-/h^+$ pair energy. In the high-energy limit ($E_r \gtrsim 50$ eV), $\epsilon_{eh}$ is constant at $\sim$ 3.6 eV \cite{Ramanathan2020}. Electrons can emit phonons via intervalley scattering, and both electrons and holes can emit phonons 
when they recombine; both processes are modeled in G4CMP \cite{Kelsey2023}. Simulated particles in Geant4 and G4CMP do not interact, therefore charges are modeled to recombine with some pre-existing partner charge. When this occurs, half of the bandgap energy (1.17~eV for Si) is emitted via phonons at the Debye frequency (15~THz for Si). These phonons initially transport diffusively and quickly transition to ballistic transport as the phonons relax in energy. This transition occurs since phonon scattering rates from isotopic scattering and anharmomic effects strongly depend on phonon energy \cite{Kelsey2023}. Because particles in G4CMP are not mutually interacting, our modeling of the $e^-/h^+$ pair burst from an impact event focuses on characterizing the phonon production and subsequent dynamics from a single $e^-/h^+$ pair of energy 3.6~eV. The band gap in Si is 1.17~eV, thus, the 2.43~eV of kinetic energy is uniformly distributed between the electron and hole. Following initialization, the electrons and holes undergo transport until they either trap on a defect in the crystal with a probability set by the trapping length $\lambda_{trap}$ or encounter a boundary of the Si substrate. We use $\lambda_{trap} = 300~\mu$m, which was determined in a study of a similar device in Ref.~\cite{Wilen2021}.

\subsection{Film Boundaries}
\label{App:Modeling_film}

The superconducting and normal metal film boundaries in our simulation perfectly absorb any $e^-$ and $h^+$ that make it to the boundary without getting trapped in the Si. However, phonons of sufficient energy can scatter and downconvert in the metal films. A pair-breaking phonon ($ E_{ph}\geq 2\Delta$, where $\Delta$ is the superconducting gap) 
entering the superconducting film from the substrate has a probability to traverse the film thickness and return to the substrate without breaking a Cooper pair given by 
\begin{equation}
P_{\textrm{escape}} = \exp\left[{-\frac{2l}{\lambda(E_{ph})}}\right],
\label{eq:prob_escape}
\end{equation}
where $\lambda$ is the energy-dependent phonon mean free path and $l$ is the distance the phonon will travel, which is twice the film thickness $d$. The factor of 2 in the numerator results from the integration over all possible angles of phonon incidence on the film. 
The phonon mean free path for both a normal metal and a superconductor is $\lambda = \nu_{\textrm{s}}/\Gamma$, where $\nu_{\textrm{s}}$ is the isotropic speed of sound in the film, $\Gamma$ is the pair-breaking rate $\Gamma^b_{ph}(E_{ph})$ for a superconducting film, or the phonon-electron relaxation rate $\Gamma^N_{\rm ph-e}$ for a normal metal film.
To calculate the phonon pair-breaking rate in a superconducting film, we use the best linear approximation over the range $2\Delta$ to $10\Delta$ for the energy dependence of the rate, which can be expressed in terms of the spectral density $S_+$ in Eq.~(90) of Ref.~\cite{Glazman2021} [cf. Eq.~(8) in Ref.~\cite{Martinis2021}]:
$\Gamma^b_{ph}(E_{ph}) = (1/\tau^{ph}_0)\left\{1+0.29\left[(E_{ph}/\Delta)-2\right] \right\}$ where $\tau^{ph}_0$ is the characteristic phonon lifetime in the superconductor \cite{Kelsey2023,Kaplan1976}. For a normal metal film, the phonon scattering rate has the form $\Gamma^N_{\rm{ph-e}} = a \omega$, where $\hbar\omega$ is the phonon energy and $a$ is a dimensionless parameter (see Appendix \hyperref[App:Sim_Params_Cu]{B.2} for details). If the phonon does undergo a pair-breaking scattering event, two QPs are generated in the superconducting film,
where the energies are given by an accept-reject loop according to Eq.~(19) of Ref.~\cite{Kelsey2023}. These QPs subsequently relax to the gap edge via phonon emission, where the resulting energies are selected via an accept-reject loop given by Eq.~(20) of Ref.~\cite{Kelsey2023}.
If either of the emitted phonons has energy $\geq 2\Delta$, they can break another pair and this cycle continues until the resulting phonons are below $2\Delta$ 
and all QPs have energy $<3\Delta$. There is a probability for the phonons to escape the film during the energy cascade, which is set by $P_{\textrm{escape}}$ from Eq.~(\ref{eq:prob_escape}) with the parameter $l$ uniformly set to $3d/2$ ($d/2$), if the phonon is directed away from (toward) the substrate, with the assumption that the scattering events occur at the center of the film. Once the QP production simulation finishes, the phonons that return to the substrate are again modeled as 3D particle tracks and the energy deposited $N_{qp}\Delta$ in the film is recorded, where $N_{qp}$ is the number of QPs generated by the scattering event. 

The simulated normal metal film follows the steps in the Monte-Carlo QP energy cascade described previously for a superconducting film, but we take the limit of $\Delta \rightarrow 0$. Thus, following the approach of Ref.~\cite{Martinis2021},  the first step in the energy cascade following the scattering of a phonon of energy $E_{ph}$ results in two excitations (an electron and a hole) whose energy is a uniform random distribution of the incident phonon energy. The generated excitations of energy below 2$\Delta_{{\rm junc}}$ are no longer tracked since they do not have enough energy to generate a pair-breaking phonon in the simulated junction material with superconducting gap $\Delta_{{\rm junc}}$. The other excitations of energy $E_e$ are modeled to relax via phonon emission, where the energy of the emitted phonon $E_{ph}$ is chosen by an accept-reject loop using the function 
\begin{equation}
 f(E_{ph}|E_{e}) = E_{ph}(E_e-E_{ph})^2,
 \label{eq:accept_rej_func_normal}
\end{equation}
which is simply taking $\Delta = 0$ in Eq.~(20) from Ref.~\cite{Kelsey2023}. The excitation then has energy $E_e-E_{ph}$. This cascade continues until all excitations and phonons in the film have energy below 2$\Delta_{{\rm junc}}$. The phonons also have a probability to escape the film and travel back into the Si, as described earlier for a superconducting film. Note that for this case, the mean free path $\lambda(E_{ph})$ in Eq.~(\ref{eq:prob_escape}) is given by the isotropic sound speed divided by the phonon-electron relaxation rate. 

\section*{Appendix B: Simulation Parameters}
\label{App:Sim_Params}
\setcounter{subsection}{0}

\begin{figure}[!t]
\centering
 \includegraphics[width= 3.4in]{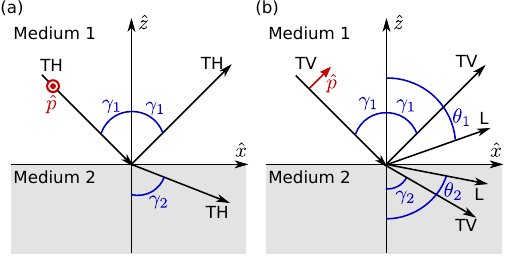}
 \caption{{\bf Transverse elastic waves at the interface of two semi-infinite media.} (a) Transverse elastic wave with its polarization perpendicular to the plane of incidence (TH) resulting in a reflected and transmitted TH wave. (b) Transverse elastic wave with the polarization is in the plane of incidence (TV) resulting in a set of reflected and transmitted Longitudinal (L) and TV waves.}
\label{fig:wavevec_transverse}
\end{figure}

G4CMP captures phonon transport physics at interfaces with transmission and either specular or diffuse reflection probabilities. In our modeling, we simulate only diffuse reflection at all boundaries. Phonons are modeled to reflect diffusively due to the rough vertical boundaries from dicing the device and the unpolished back-side surface. The absorption probability $p_{abs}$ is the probability of transmission from Si to a device-layer film. In general, this depends on both the polarization and angle of incidence of the phonon. We can represent the phonons as elastic plane waves incident on an interface of two semi-infinite media, which is in close analogy to electromagnetic waves incident on an interface. However, elastic plane waves differ in that a longitudinal polarization is allowed. The simplest case is the transverse horizontal polarization (TH), diagrammed in Fig.~\ref{fig:wavevec_transverse}(a), where the incident wave polarization direction $\hat{p}$ is perpendicular to the plane of incidence (the $\hat{x}-\hat{z}$ plane in Fig.~\ref{fig:wavevec_transverse}). Due to the continuity conditions of the displacement vectors and the material stress tensor in both the normal and tangential directions to the interface of the media, the displacement vectors of the waves produced from this incident wave can only be in the same direction as the incident wave \cite{Kaplan1979}. Therefore, only TH waves are produced. The situation becomes more complicated when the incident wave displacement vector has a non-zero component along the normal direction $\hat{z}$ of the interface. This is the case for the transverse vertical polarization (TV), shown in Fig.~\ref{fig:wavevec_transverse}(b). If the incident wave displacement vectors have components parallel and perpendicular to the interface normal $\hat{z}$, both a longitudinal (L) and TV wave can be produced from the incident wave, and as a result, four waves are produced from the scattering site. A third case occurs when the incident wave has a longitudinal polarization, where the displacement vectors are in the same direction as the wavevector. The scattering of such a wave can also produce four waves for similar reasons as in the TV case, where there exists a set of longitudinal and transverse vertical waves.

In G4CMP we set simple absorption/reflection probabilities for the phonons when they scatter off the device films. We follow Ref.~\cite{Kaplan1979} and compute the angle-averaged transmission probabilities for incident transverse horizontal, transverse vertical, and longitudinal waves \cite{Wei2022}. For incident transverse phonons, we consider the polarization direction relative to the plane of incidence to be randomly distributed between 0 and $\pi$/2. As a result, the transmission probability of transverse waves $\eta_T$ is the arithmetic mean of the angle-averaged TV and TH cases \cite{Kaplan1979}. We then take the absorption probability from the Si substrate into a film to be the sum of the angle-averaged transmission probabilities ($\eta_T$ and $\eta_L$) weighted by the density of states for the phonons within the substrate
\begin{equation}
p_{abs} = \textrm{DOS}^{\textrm{Si}}_{T}\eta_T+\textrm{DOS}^{\textrm{Si}}_{L}\eta_L,
\end{equation}
where $\textrm{DOS}^{\textrm{Si}}_{T}~(\textrm{DOS}^{\textrm{Si}}_{L})$ is the density of states of transverse (longitudinal) modes in Si and $\eta_T~(\eta_L)$ is the transverse (longitudinal) angle-averaged transmission probability from Si into the film. Note that the above calculation approximates the two
media as isotropic and assumes the interface between the two media to be perfectly
smooth. This calculation is an approximation for the reflection/absorption simulation parameters at the interfaces and is not intended to be a detailed physical
description of the phonon scattering at arbitrary interfaces, which is beyond the
scope of our present work.


We use a similar scheme to define the isotropic speed of sound $\nu_s$ in defining the phonon mean free path in both superconducting and normal metal films. We weight the sound speeds of the various modes by the density of phonon polarizations within the Si: 
\begin{equation}
\nu_s = \textrm{DOS}^{\textrm{Si}}_{T}\nu_T + \textrm{DOS}^{\textrm{Si}}_{L}\nu_L,
\label{eq:v_sound}
\end{equation}
where 
$\nu_T~(\nu_L)$ is the transverse (longitudinal) mode sound speed in the material. See Table \ref{tab:simulation parameters} for values of simulation parameters used to model device films.

\begin{table}[!htbp]
    \centering
    \begin{tabular}{ |p{2.5cm}||p{1.cm}|p{1.2cm}|p{1.cm}|p{1.cm}| }\hline
    \multicolumn{5}{|c|}{\makecell[b]{Simulation Material Parameters}}\\\hline
     Parameter & \multicolumn{4}{c|}{Material}  \\\hline
          & \hfil Al & \hfil Nb & \hfil Ti & \hfil Cu\\\hline
        $\rho~(\textrm{g}~\textrm{cm}^{-3})$~\cite{Kittel2005} & \hfil2.7 & \hfil8.58 & \hfil4.5 &\hfil 8.93 \\\hline
        $\nu_{T}~(\mu\textrm{m}~\textrm{ns}^{-1})$~\cite{SimmonsWang} &\hfil 3.251 & \hfil2.168 & \hfil3.33 &\hfil 2.38 \\\hline
        $\nu_{L}~(\mu\textrm{m}~\textrm{ns}^{-1})$~\cite{SimmonsWang} &\hfil 6.808 & \hfil5.139 &\hfil 6.254 &\hfil 4.828 \\\hline
        $\nu_{s}~(\mu\textrm{m}~\textrm{ns}^{-1})$ & \hfil3.58 &\hfil 2.44 &\hfil 3.60 & \hfil2.61 \\\hline
        $\eta_{T}$~\cite{Kaplan1979,Wei2022} &\hfil 0.776 & \hfil 0.736 & \hfil 0.776 &\hfil 0.727 \\\hline
        $\eta_{L}$~\cite{Kaplan1979,Wei2022} &\hfil 0.98 & \hfil 0.835 &\hfil 0.939 &\hfil 0.826 \\\hline
        $p_{abs}$ &\hfil 0.795 &\hfil 0.745 &\hfil 0.792 &\hfil 0.736 \\\hline
        $\Delta~(\mu\textrm{eV})^{\textrm{a}}$ & \hfil180$^{\textrm{c}}$ & \hfil1538 & \hfil59 &\hfil - \\\hline
        $\tau_0^{ph} ~(\textrm{ns})^{\textrm{b}}$ &\hfil 0.242 & \hfil0.00417 & \hfil0.414 &\hfil \hyperref[App:Sim_Params_Cu]{B.2} \\\hline
    \end{tabular}
    \caption{Simulation parameters.
    \newline $^{\textrm{a}}$ Values are calculated from Table IV of Ref.~\cite{Carbotte1990}. \newline $^{\textrm{b}}$ Values for Al and Nb are from Table II of Ref.~\cite{Kaplan1976}. See discussion in Appendix~\hyperref[App:Sim_Params]{B} for values for Ti and Cu. \newline $^{\textrm{c}}$ In general, the gap of Al is simulated to be thickness dependent \cite{Marchegiani2022}.}
    \label{tab:simulation parameters}
\end{table}

\subsection{Phonon lifetime $\tau^{ph}_0$ in Ti}

For the superconducting films modeled in this work, we use the phonon lifetime from Table II of Ref.~\cite{Kaplan1976} (see Table \ref{tab:simulation parameters}). The phonon lifetime is defined to be 
\begin{equation}
\tau^{ph}_0 = \frac{\hbar N}{4\pi^2N(0)\langle \alpha^2 \rangle_{\textrm{av}}\Delta},
\label{eq:tau_0}
\end{equation}
where $N(0)$ is the single-spin density of states at the Fermi energy, $N$ is the atomic density of the material, and $\Delta$ is the superconducting gap at $T=0$. Note that $\langle \alpha^2 \rangle_{\textrm{av}}$ is given by 
\begin{equation}
3\langle \alpha^2 \rangle_{\textrm{av}} = \int^{\infty}_0\alpha^2(\Omega)F(\Omega)\textrm{d}\Omega,
\label{eq:alpha_av}
\end{equation}
where $\alpha(\Omega)$ is the matrix element of the electron-phonon interaction and $F(\Omega)$ is the phonon density of states. The square of the matrix element $\alpha^2(\Omega)$ can be extracted from measurements of $\alpha^2(\Omega)F(\Omega)$ via tunneling experiments and then dividing out $F(\Omega)$ from neutron scattering data \cite{Kaplan1976,McMillan1969Parks}. To our knowledge, there is no published tunneling data for Ti, but there does exist neutron scattering data \cite{Stassis1979}. 

For the case of Nb, in Table II of Ref.~\cite{Kaplan1976} tunneling data was not used and $\alpha^2(\Omega)$ was approximated to be constant and its value was estimated with the mass enhancement factor $\lambda = 1.84$. The mass enhancement factor is defined as 
\begin{equation}
\lambda = \int_0^{\infty}2\frac{\alpha^2F(\Omega)\textrm{d}\Omega}{\Omega}, 
\label{eq:lambda}
\end{equation}
which is twice the inverse moment of $\alpha^2F(\Omega)$ \cite{Carbotte1990}. 

For Ti, we also assume $\alpha^2$ to be constant so that we can rewrite Eq.~(\ref{eq:lambda}) for $\alpha^2$ as a function of $\lambda$ and the phonon density of states $F(\Omega)$.
Using numerical integration of the phonon density of states $F(\Omega)$ data from Ref.~\cite{Stassis1979} and a value of $\lambda = 0.38$ from Table III of Ref.~\cite{McMillan1968}, we find that $\langle \alpha^2 \rangle_{\textrm{av}} = 1.3~\textrm{meV}$. We then use values of the single-spin electron density of states $N(0)$ and the atomic density $N$ from Table VI in Ref.~\cite{Gladstone1969Parks}. We assume Ti to be a weakly-coupled superconductor, with $\Delta = 1.76 k T_c$ and $T_c = 0.39~\textrm{K}$ from Table VI in Ref.~\cite{Gladstone1969Parks}. We then arrive at a value of $\tau^{ph}_0 = 0.414~\textrm{ns}$ for the phonon lifetime of Ti, which is roughly a factor of two longer than the phonon lifetime in Al \cite{Kaplan1976}.

\subsection{$\Gamma^{N}_{\rm ph-e}$ for Cu}
\label{App:Sim_Params_Cu}

The electroplated Cu films on our experimental devices are relatively clean, with RRR $\sim$42 \cite{Iaia2022}. Therefore, to estimate the phonon lifetime for our simulations, we assume a pure Cu film and use the Debye model.
In a clean conductor, primarily longitudinal phonons scatter from electrons. The decay rate for a longitudinal phonon in a clean normal conductor in the low-temperature limit for a phonon of energy $\hbar \omega$ can be written as 
\begin{equation}
    \Gamma^N_{{\rm ph-e}} = \pi \beta_L \frac{\nu_L}{\nu_F} \omega,
    \label{eq:normal_phonon_rate}
\end{equation}
where $\nu_F$ is the Fermi velocity, $\beta_L$ is a dimensionless coupling constant given by $(2\epsilon_F/3)^2N(0)/2\rho \nu^2_L$, $\epsilon_F$ is the Fermi energy \cite{Sergeev_2000}.
From Ref.~\cite{AshcroftMermin}, Cu has the following values: $\epsilon_F = $ 7 eV, $\nu_F = $ 1.57$\times10^{6}$~m/s, and $N(0)$ =  1.12$\times10^{47}$~J$^{-1}$m$^{-3}$. With this, we arrive at $\beta_L$ = 0.16, and $1/\Gamma^{N}_{\rm{ph-e}}$ = 1.2 ns for a phonon of energy 2$\Delta_{{\rm Al}}$ in Cu.

\section*{Appendix C: Experimental device parameters}
\label{App:Exp_Params}
\setcounter{subsection}{0}

  Qubit parameters for the non-Cu-B and the 1-$\mu$m Cu devices can be found in Table~\ref{tab:exp_param}. Similar data for the non-Cu-A and the 10-$\mu$m Cu devices are reported in Ref.~\cite{Iaia2022}. The resistance $R_n$ for the injector junction  used to inject phonons for the measurements shown in Fig.~\ref{fig:injection_Nb_1umCu} and Fig.~\ref{fig:injection_Nb_1umCu_AllQubits} in the non-Cu-B (1-$\mu$m Cu) device is 6.9~k$\Omega$ (5.5~k$\Omega$).

\begin{table}[!htbp]
\begin{tabular}{ |p{1.5cm}||p{1cm}|p{1.5cm}|p{1.5cm}|p{1.5cm}|}
 \hline
 \multicolumn{5}{|c|}{Qubit Parameters} \\ \hline
 
 \hfil Device & \hfil Qubit & \hfil $f_{01} {\rm (GHz)}$ & \hfil $T_{1}{\rm (}\mu{\rm s)}$ & \hfil $\delta f{\rm(MHz)}$ \\
 \hline
 

 \hfil\multirow{6}*{Non-Cu-B}&  \hfil $Q_1$ & \hfil 4.60 & \hfil 44(12)  &  \hfil 3.075 \\
                          &  \hfil $Q_2$ & \hfil 4.58 & \hfil 31(6) &  \hfil 1.80 \\
                          &  \hfil $Q_3$ & \hfil 4.11  & \hfil 22(2)&  \hfil 4.95\\
                          &  \hfil $Q_4$ & \hfil 4.32   & \hfil 26(3)&   \hfil 5.80\\
                          &  \hfil $Q_5$ & \hfil 4.10  & \hfil 10(1)&  \hfil 8.00\\
                          &  \hfil $Q_6$ & \hfil 4.18  & \hfil 44(4) &   \hfil  4.44\\
 \hline

 \hfil\multirow{6}*{1-$\mu$m Cu}&  \hfil $Q_1$ & \hfil 4.62 & \hfil 18(3) & \hfil 2.78 \\
                          &  \hfil $Q_2$ & \hfil 4.30 & \hfil 24(2) &  \hfil 14.8 \\
                          &  \hfil $Q_3$ & \hfil 4.32  & \hfil 24(4) &  \hfil 3.85\\
                          &  \hfil $Q_4$ & \hfil 4.36 & \hfil 17(3) &  \hfil 5.77\\
                          &  \hfil $Q_5$ & \hfil 4.34 & \hfil 19(2) &  \hfil 5.40\\
                          &  \hfil $Q_6$ & \hfil 4.32  & \hfil 22(3) &  \hfil 3.51\\
\hline
\end{tabular}
\caption{Qubit parameters for the non-Cu-B and 1-$\mu$m Cu devices. Note that similar data for the other devices (non-Cu-A and 10-$\mu$m Cu) is reported in Ref.~\cite{Iaia2022}.}
\label{tab:exp_param}
\end{table}

\section*{Appendix D: Injection Experiments}
\label{App:Exp_Inj}
\setcounter{subsection}{0}

\begin{figure*}[t!]
\centering
 \includegraphics[width=6.8in]{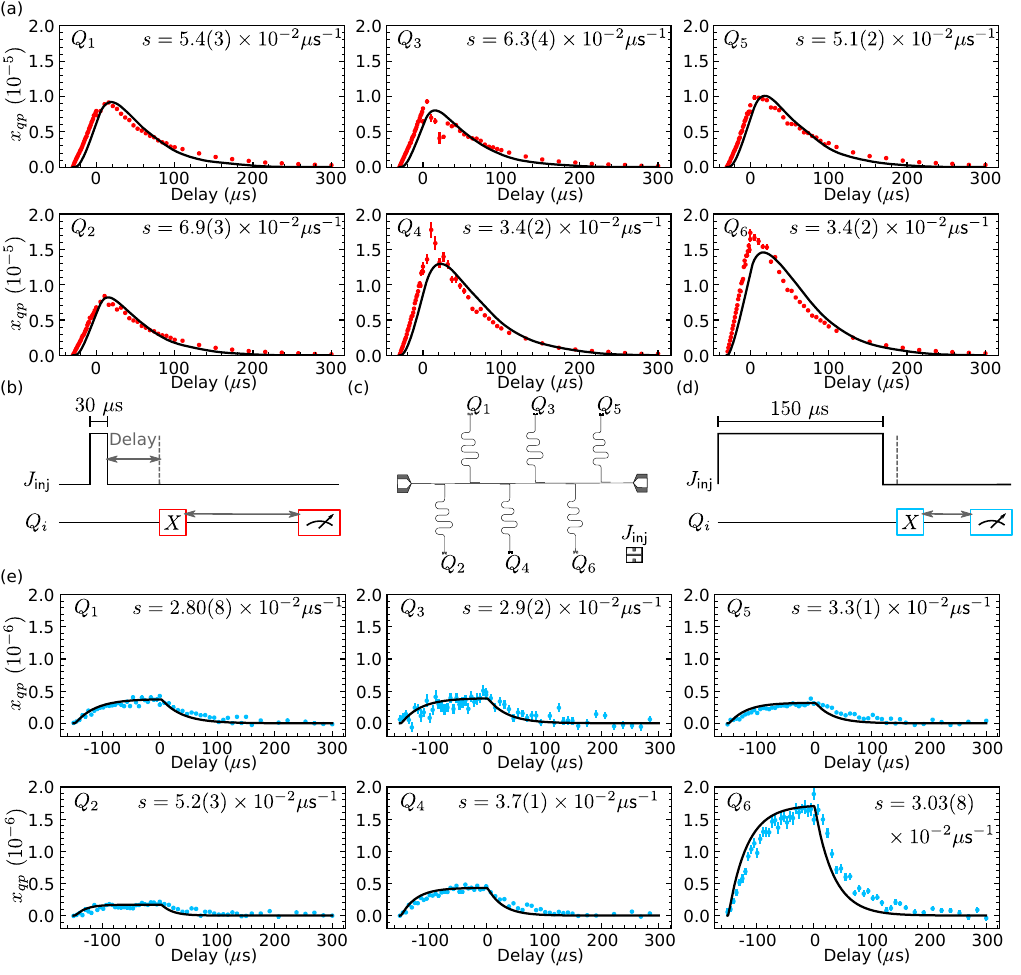}
 \caption{{\bf Experimental and simulated QP response to phonon injection for all qubits on the non-Cu-B and 1-$\mu$m Cu devices.} 
 (a) Measured (red) and simulated (black) $x_{qp}(t)$ for all six qubits on the non-Cu-B device for a 30-$\mu$s pulse. The resulting trapping rates from fitting the simulation curve to the data are in each figure. (b) Pulse sequence for the non-Cu-B injection. (c) Physical location of qubits relative to the injector junction $J_{\sf inj}$. (d) Pulse sequence for phonon injection in the 1-$\mu$m Cu device with a 150-$\mu$s injection pulse. (e) The measured (light blue) and simulated (black) $x_{qp}(t)$ from the 150-$\mu$s pulse in the 1-$\mu$m Cu device. Note different scales for the $x_{qp}$ axis in (a) and (e).
 \label{fig:injection_Nb_1umCu_AllQubits}}
\end{figure*}

We perform controlled phonon injection experiments on all six qubits on the two new experimental devices for this work, where we previously focused on measuring only three qubits in Ref.~\cite{Iaia2022}. This allows for a detailed study of the $\Delta \Gamma_1$ response versus distance from the phonon injection location. We find that the qubit closest to the injection location ($Q_6$) experiences the highest level of $x_{qp}$ relative to the other qubits in the array for both devices. However, we need a much longer injection pulse to see a measurable effect on $\Delta\Gamma_1$ for the 1-$\mu$m Cu chip compared to the device without back-side metallization (non-Cu-B). In Fig.~\ref{fig:injection_Nb_1umCu_AllQubits} we see the measured and simulated $x_{qp}(t)$ response to an injection pulse for all six qubits on both the non-Cu-B and 1-$\mu$m Cu chips. From our modeling of our experiments, we arrive at trapping rates $s$ that are comparable to measurements of other planar multi-layer qubits \cite{Leonard2019,Liu2023,KLiu2023,Bargerbos2023}. We note that slower trapping rates have been observed for single-layer, all-Al qubits measured in 3D cavities \cite{Wang2014,Diamond2022}.

Our measurement sequence of the increase in qubit relaxation rate $\Delta \Gamma_1 = 1/T_1 - 1/T^b_1$ under controlled phonon injection involves biasing an on-chip tunnel junction with a square pulse of length $T_{\rm pulse}$ and amplitude 1~mV. After this pulse, we measure the qubit $T_1$ via a sequence of idle times after an $X$ pulse. We interleave measurements without biasing the on-chip junction to determine the baseline coherence time $T^b_1$. In this work, we present values of $\Delta\Gamma_1$ at negative delay values, corresponding to $T_1$ measurements during the pulse. One potential issue arises because our measurements of $\Delta\Gamma_1$ are not instantaneous in time, since for each $T_1$ we measure a range of delays after the $X$ pulse up to 100~$\mu$s. This means that the measured $T_1$ values are sensitive to excess QP density during this time range. To counteract this effect, for the $T_1$ measurements during the pulse, we end the injection pulse when we apply the $X$ pulse to excite the qubit for the $T_1$ measurement. We 
average 20 rounds of this experiment so that our data is not sensitive to fluctuations in the baseline $T^b_1$.

\subsection{$x_{qp}$ fit parameters}
\label{App:Exp_Inj_sim_params}

\begin{figure}[!t]
\centering
 \includegraphics[width= 3.4in]{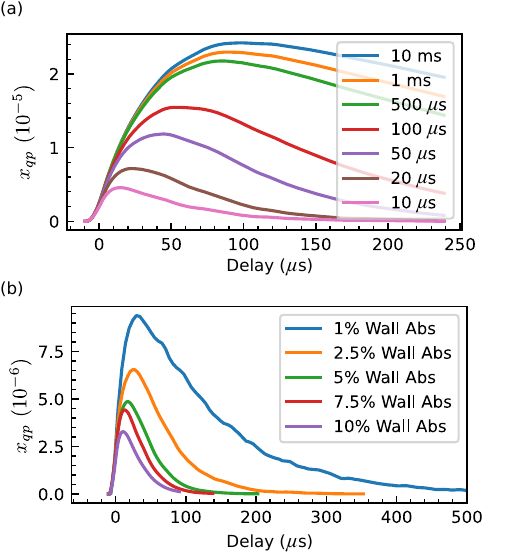}
 \caption{{\bf Dependence of injection response on QP trapping and phonon loss.} Example of simulated $x_{qp}(t)$ curves for $Q_4$ on the non-Cu-A chip (see Fig.~\ref{fig:phonon-injection}). (a) We vary the value of the trapping time $s^{-1}$ while keeping the wall absorption constant at 2.5$\%$. (b) We vary the wall absorption probability in various G4CMP runs while keeping the trapping time $s^{-1}$ constant at 20~$\mu$s.}
\label{fig:Trap_Wall_Abs}
\end{figure}

When simulating the $x_{qp}$ response for each qubit during controlled phonon injection, we introduce two free parameters: the wall absorption probability and the QP trapping rate $s$. The inclusion of the wall absorption probability accounts for the escape of phonons from the substrate via Al wire-bonds that surround the top-side of the device layer, GE-varnish that adheres the device to the machined Al box, and scattering from the rough vertical boundaries of the device due to dicing of the wafer into chips. 
We find that a probability of 2.5$\%$ best captures the shape of our measured $x_{qp}(t)$ response following an injection pulse. Because the QP trapping rate depends on microscopic details in a given physical device, we set this to be a free parameter when fitting the measured $x_{qp}(t)$ response. These two parameters influence the shape of our modeled $x_{qp}(t)$ response in different ways (see Figure~\ref{fig:Trap_Wall_Abs}). The value of $s$ influences the saturation level ($\sim g/s$) and the characteristic recovery time. A reduction in the wall absorption probability increases the phonon lifetime in the substrate, which in turn increases the overall number of phonons incident at the qubit junction. However, for simulated devices with a down-converting film on the back side, the $x_{qp}$ response is relatively insensitive to the wall absorption probability since the phonons are being downconverted much faster than they leave the substrate via the vertical boundaries. 

As described in the main text, our modeling of the phonon injection experiments based on Eq.~(\ref{eq:xqp_model}) is insensitive to QP recombination due to the relatively small $x_{qp}$ levels. 
For our devices, we find a typical trapping rate to be $5\times10^{-2}\mu{\rm s}^{-1}$. 
For a recombination rate of $1/(10~{\rm ns})$ based on prior measurements in the literature \cite{Kaplan1976,Ullom1998,Nsanzineza2014,Wang2014}, the trapping and recombination terms in Eq.~(\ref{eq:xqp_model}) become comparable at $x_{qp}\sim5\times10^{-4}$.
Therefore, trapping is the dominant process when modeling our injection experiments, where the highest $x_{qp}$ level we measure is $\sim1.8\times10^{-5}$. For the modeling of the typical $\gamma$-ray impacts, a qubit directly above the impact location experiences a peak in $x_{qp}$ at $6\times10^{-4}$ and from Fig.~\ref{fig:Edge_center} we can see that the qubit close to the impact location [position (i)] reaches a maximum $x_{qp}$ of $5.6\times10^{-5}$. 
Thus, recombination is comparable to trapping for qubits closest to the impact site for short times after the impact, although trapping is the dominant process for most qubits in the simulated array away from the impact. 
Therefore, we include the recombination term in our modeling of both the phonon injection and $\gamma$ impact processes, although it only plays a role for a small fraction of the qubits.

\section*{Appendix E: Details of QP charge-parity switching measurements}
\label{App:Exp_ChargeParity}
\setcounter{subsection}{0}

 In the main paper, we present new experimental results for devices with a six-qubit array of charge-sensitive transmons: the non-Cu-B and 1-$\mu$m Cu chips. To compare the new devices to those described in Ref.~\cite{Iaia2022}, we use identical measurement sequences described in that prior work. We utilize the charge dispersion $\delta f$ of our transmons to probe the charge-parity state of our qubit islands as a measure of QP poisoning in our qubits. We use a modified Ramsey sequence consisting of an $X/2$ pulse, idle for $1/4\delta f$, and finally a $Y/2$ pulse that maps the even and odd parity states to either the 0 or 1 state \cite{Riste2013, Serniak2018, Christensen2019, Iaia2022}. We repeat this sequence 20,000 times with a repetition period of 10~ms. We create a digital time trace of this data using a simple threshold technique and then compute a power spectral density (PSD) of the resulting digital signal. After averaging PSDs from multiple 20,000 single-shot experiments, we fit a Lorentzian to the spectrum, as done in Ref.~\cite{Riste2013}
\begin{equation}
S_{\rm p}(f)=\frac{4F^2\Gamma_{\rm p}}{(2\Gamma_{\rm p})^2+(2\pi f)^2}+\left( 1-F^2\right)\Delta t,
\end{equation}
where $\Gamma_{\rm p}$ is the characteristic parity switching rate, $F$ is the parity sequence mapping fidelity, and $\Delta t$ is the repetition period of the measurement. For all six qubits on the non-Cu-B device, we found an average $\Gamma_{\rm p}$ of 0.6(1)~s$^{-1}$, which is within a factor of two of the non-Cu-A data from Ref.~\cite{Iaia2022}.

For detecting simultaneous charge-parity switching events, we follow the data analysis routine done in Ref.~\cite{Iaia2022}. We repeatedly measure the background charge-parity for $\sim$16.5 hrs for the 1-$\mu$m Cu device with a repetition rate of 10~ms. We use a Hidden Markov model (HMM) to digitize the parity switching data. As in Ref.~\cite{Iaia2022}, we define simultaneous parity switches to occur when edges of the digitized data fall in the same window $\Delta t_w=400$~ms. Note that this window of 400~ms is much smaller than the shortest charge-parity lifetime of $\sim$2.7~s (non-Cu-A, with no back-side metallization). The probability of a random coincident parity switch occurring on both qubit $i$ and qubit $j$ within $\Delta t_w$ is given by $(r_i\Delta t_w)(r_j \Delta t_w)$, where $r_i$ is the parity switching rate of qubit $i$. Using this, we define the random background switching rate to be the probability of a two-fold coincident parity switch randomly occurring in our detection window divided by the duration of the detection window $(r_i\Delta t_w)(r_j \Delta t_w)/\Delta t_w$. The extracted poisoning rates (open bars) in Fig.~\ref{fig:coincident-parity}(a) were computed as done in Ref.~\cite{Iaia2022}, however, the inverted set of equations is much simpler for analyzing simultaneous switching on a pair of qubits compared to a group of three qubits: 

\begin{align}
\begin{aligned}
p_A^{\rm obs} & = \frac{1}{2}\left(p_{AB}+p_A \right) \\
p_B^{\rm obs} & = \frac{1}{2}\left(p_{AB}+p_B \right) \\
p_{AB}^{\rm obs} & = \frac{1}{4}\left(p_{AB}+p_Ap_B \right). \\
\end{aligned}
\label{prob_eqs}
\end{align}

\noindent Note that the factors of $1/2$ and $1/4$ account for the fact that we can only detect an odd number of changes in the charge-parity state between measurements. Thus, on average we observe half of the parity-switching events for a single qubit. 


\section*{Appendix F: Phonon Caustics}
\label{App:Caustic}
\setcounter{subsection}{0}

\begin{figure}[!t]
\centering
 \includegraphics[width= 3.4in]{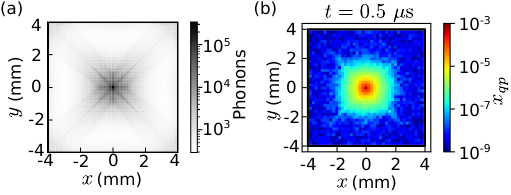}
 \caption{{\bf Example of structure from phonon caustics.} (a) Spatial distribution of phonon absorption locations on the top-side surface from a simulated burst of 20 million phonons initialized at the center of the top-side surface of the substrate with a random downward angle. (b) The $x_{qp}$ poisoning distribution in a simulated device with 10-$\mu$m thick Cu back-side islands at $t=0.5~\mu$s after a $\gamma$ impact at the center of the chip.}
\label{fig:caustic}
\end{figure}
\begin{figure*}[!t]
\centering
 \includegraphics[width= 6.8in]{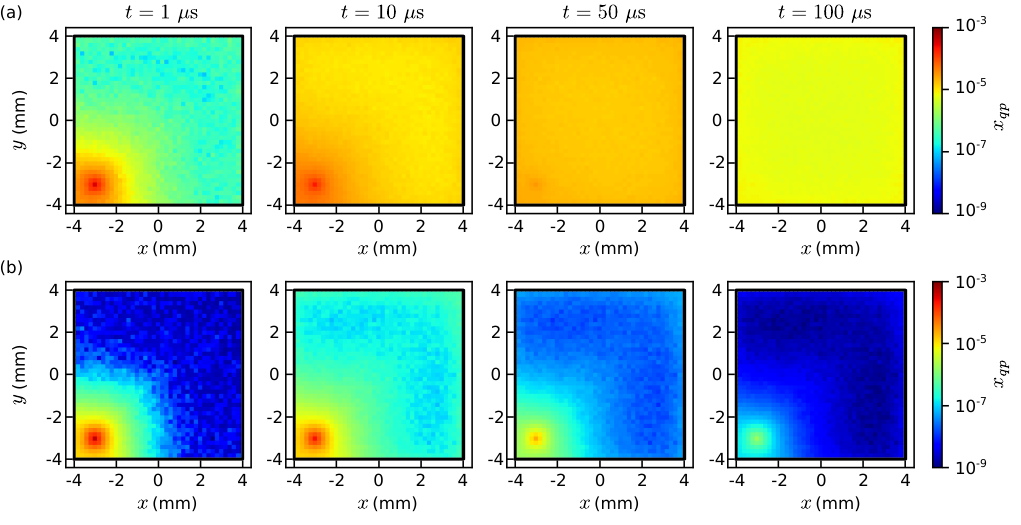}
 \caption{{\bf QP poisoning distribution following off-center $\gamma$ impact.} Modeled $x_{qp}(x,y,t)$ in qubit array identical to one shown in Fig.~\ref{fig:footprint-density-plot} with the impact site at $x=y=-3$~mm. (a) Device with no back-side metallization. (b) Device with 10-$\mu$m-thick Cu islands on back side of chip.
\label{fig:footprint-density-plot-offcenter}}
\end{figure*}

The modeling of phonon transport within G4CMP only supports the ballistic transport of acoustic phonons. The direction and magnitude of the velocity of these particles is given by the phonon group velocity $\vec{v}_g= \nabla_k \omega (\vec{k})$, where $\omega$ is the angular frequency and $\vec{k}$ is the wavevector. In general, the dependence of the phonon's angular frequency on the wavevector comes from the wave equation that governs phonon transport:
\begin{equation}
    \rho \omega^2 e_i = C_{ijml}k_jk_me_l,
    \label{eq:phonon_wave_eq}
\end{equation}
\noindent where $\rho$ is the mass density, $\vec{e}$ is the polarization vector, and $C_{ijml}$ is the elasticity tensor~\cite{Kelsey2023}. Due to anisotropy in the elasticity tensor $C_{ijml}$, the group velocity is generally not in the same direction as the crystal momentum $\hbar\vec{k}$. Thus, phonons initialized with a uniform distribution in $\vec{k}$ space will not result in a uniform distribution of ballistic phonon group velocities. The resulting structure in the spatial distribution of the phonon directions can be seen in Fig.~\ref{fig:caustic}(a) where we initialize 20 million phonons at the center of the top-side surface of the Si substrate with a random downward direction for $\vec{k}$. In this case, the top-side film is set to be perfectly absorbing. Figure~\ref{fig:caustic}(a) shows the distribution of phonons absorbed on the top-side surface and agrees with the phonon caustic structure described in Ref.~\cite{Martinez2019,Kelsey2023,Northrop1979}. We see similar structure in the $x_{qp}$ response in the modeling of a dense qubit array following a $\gamma$ impact, where in Fig.~\ref{fig:caustic}(b) we plot the $x_{qp}$ response in the simulated junctions at 500~ns after a $\gamma$ impact of 100~keV. Phonons are modeled to reflect diffusively due to the rough vertical boundaries from dicing the chip and the unpolished back-side surface. Thus, a phonon will have a random direction in $\vec{k}$ after each reflection. Therefore, we expect the phonon distribution to most closely reflect the caustic structure at early times after a particle impact. Such structure in the QP poisoning pattern due to the phonon caustics may be an important consideration in designing future layouts of dense qubit arrays.

\section*{Appendix G: Modeling off-center particle impacts}
\label{App:Off_Center}
\setcounter{subsection}{0}



The $x_{qp}$ density plots shown in Fig.~\ref{fig:footprint-density-plot} are due to a simulated particle impact at the center of the device. In general, we would expect each impact to occur at a random location within the device substrate. It is important to know if the shape of the $x_{qp}$ response varies depending on the position of the impact. To test this, we simulate a particle impact at the $x=y=-3$~mm location shown in Fig.~\ref{fig:footprint-density-plot-offcenter}. We observe a similar spatial and temporal $x_{qp}$ response at this off-center impact location compared to an impact at the center, as in Fig.~\ref{fig:footprint-density-plot}. In this case, the vertical wall boundaries are much closer to the impact location, but this proximity does not appear to affect the resulting $x_{qp}$ distribution.

\section*{Appendix H: QP Diffusion}
\label{App:Diffusion}
\setcounter{subsection}{0}

In this work, we have neglected the possibility of QP diffusion within the ground plane. Because all the devices in this work have a Nb ground plane, no QPs will be produced in the ground plane for 
injected phonons of energy $2\Delta_{{\rm Al}}$. 
However, the phonons generated from a $\gamma$ impact do have sufficient energy to break pairs in the Nb ground plane, thus we must justify leaving out QP diffusion more carefully for these simulations. In general, the QP diffusion constant depends on the QP energy $E_{qp}$: 
\begin{equation}\label{eq:Diffusion}
    D(E_{qp}) = D_n\sqrt{1-\bigg(\frac{\Delta}{E_{qp}}\bigg)^2},
\end{equation}
where $D_n$ is the normal state electronic diffusion constant, and $\Delta$ is the superconducting gap energy \cite{Ullom1998}. For large QP energies, the diffusion constant approaches the normal state diffusion, and for QPs at the gap edge the diffusion constant approaches zero. In a superconductor, energetic QPs relax to the gap edge via phonon emission. The rate for this process depends on the characteristic electron-phonon time $\tau^{qp}_0$. An approximate form of this QP relaxation rate is presented in Eq.~(5) of Ref.~\cite{Martinis2021} for a weakly coupled superconductor: $\Gamma^s_{qp} \sim (1.8/\tau^{qp}_0)\left[(E_{qp}/\Delta)-1\right]^3$ \ (an exact formula for this case can be found in Appendix~B of Ref.~\cite{Marchegiani2022}). We define a characteristic QP lifetime as $1/\Gamma^s_{qp}$. Thus, we define an energy-dependent QP diffusion length as $\sqrt{D(E_{qp})/\Gamma^s_{qp}(E_{qp})}$. This diffusion length depends only on QP energy and material parameters. We measured the resistivity of our fabricated Nb films at $\sim$10~K to be 6.07~$\mu\Omega\cdot$cm, which results in a normal-state diffusion constant of 0.88~$\mu$m$^2$ns$^{-1}$. This diffusion constant is roughly an order of magnitude lower than for an Al film with a normal state diffusion constant of 6~$\mu$m$^2$ns$^{-1}$ \cite{Ullom1998,Friedrich1997}. Similarly, the electron-phonon time for Nb is 0.15~ns, which is three orders of magnitude lower than the time for Al of 440~ns. Both of these material parameters result in a significantly lower characteristic QP diffusion length in a Nb film as compared to an Al film.

\begin{figure}[!b]
\centering
 \includegraphics[width= 3.4in]{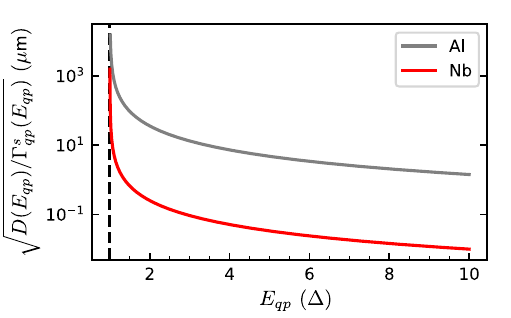}
 \caption{{\bf Characteristic diffusion length.} Comparison of characteristic diffusion lengths for both Al and Nb films versus QP energy in units of the respective superconducting gap $\Delta$ for each material.}
\label{fig:diff_length}
\end{figure}

In Fig.~\ref{fig:diff_length} we plot the characteristic diffusion length for both a Nb and Al film as a function of QP energy $E_{qp}$ (in units of the gap $\Delta$ for each material). For both cases, as $E_{qp}$ approaches $\Delta$, the characteristic diffusion length diverges; although the diffusion constant $D$, Eq.~\eqref{eq:Diffusion}, goes to zero, the relaxation rate $\Gamma^s_{qp}$ vanishes faster than $D$. However, this apparent divergence is eventually cut by, for instance, finite temperature effects (as QPs can be excited by absorbing thermal phonons), the possible recombination between two QPs, or mechanisms causing broadening of the superconducting density of states. The plot shows that in Al, QPs with energy up to several $\Delta$ diffuse over lengths of at least a few $\mu$m; therefore, treating our $5\,\mu$m-long Al junction electrodes as zero-dimensional (that is, hosting a uniform QP density) is a good approximation. For Nb, the diffusion length remains below $10\,\mu$m down to energies of order $1.06\Delta$ (note that for Nb, density of state broadenings on the order of several percent have been reported in the literature~\cite{Nevala2012,Julin2016}); even for the simulations with dense arrays of qubits, their distance is much longer than $10\,\mu$m, so we expect diffusion not to play a significant role. Conversely, in an Al groundplane QPs with energy below $1.28\Delta$ would diffuse over the interqubit distance of $200\,\mu$m, and hence it may not be possible to ignore diffusion in such a case.

\bibliography{QPRefs_92121}

\begin{thebibliography}{56}
\expandafter\ifx\csname natexlab\endcsname\relax\def\natexlab#1{#1}\fi
\expandafter\ifx\csname bibnamefont\endcsname\relax
  \def\bibnamefont#1{#1}\fi
\expandafter\ifx\csname bibfnamefont\endcsname\relax
  \def\bibfnamefont#1{#1}\fi
\expandafter\ifx\csname citenamefont\endcsname\relax
  \def\citenamefont#1{#1}\fi
\expandafter\ifx\csname url\endcsname\relax
  \def\url#1{\texttt{#1}}\fi
\expandafter\ifx\csname urlprefix\endcsname\relax\def\urlprefix{URL }\fi
\providecommand{\bibinfo}[2]{#2}
\providecommand{\eprint}[2][]{\url{#2}}

\bibitem[{\citenamefont{McEwen et~al.}(2022)\citenamefont{McEwen, Faoro, Arya,
  Dunsworth, Huang, Kim, Burkett, Fowler, Arute, Bardin et~al.}}]{McEwen2021}
\bibinfo{author}{\bibfnamefont{M.}~\bibnamefont{McEwen}},
  \bibinfo{author}{\bibfnamefont{L.}~\bibnamefont{Faoro}},
  \bibinfo{author}{\bibfnamefont{K.}~\bibnamefont{Arya}},
  \bibinfo{author}{\bibfnamefont{A.}~\bibnamefont{Dunsworth}},
  \bibinfo{author}{\bibfnamefont{T.}~\bibnamefont{Huang}},
  \bibinfo{author}{\bibfnamefont{S.}~\bibnamefont{Kim}},
  \bibinfo{author}{\bibfnamefont{B.}~\bibnamefont{Burkett}},
  \bibinfo{author}{\bibfnamefont{A.}~\bibnamefont{Fowler}},
  \bibinfo{author}{\bibfnamefont{F.}~\bibnamefont{Arute}},
  \bibinfo{author}{\bibfnamefont{J.~C.} \bibnamefont{Bardin}},
  \bibnamefont{et~al.}, \bibinfo{journal}{Nature Physics}
  \textbf{\bibinfo{volume}{18}}, \bibinfo{pages}{107} (\bibinfo{year}{2022}).

\bibitem[{\citenamefont{Veps{\"a}l{\"a}inen
  et~al.}(2020)\citenamefont{Veps{\"a}l{\"a}inen, Karamlou, Orrell, Dogra,
  Loer, Vasconcelos, Kim, Melville, Niedzielski, Yoder
  et~al.}}]{Vepsalainen2020}
\bibinfo{author}{\bibfnamefont{A.~P.} \bibnamefont{Veps{\"a}l{\"a}inen}},
  \bibinfo{author}{\bibfnamefont{A.~H.} \bibnamefont{Karamlou}},
  \bibinfo{author}{\bibfnamefont{J.~L.} \bibnamefont{Orrell}},
  \bibinfo{author}{\bibfnamefont{A.~S.} \bibnamefont{Dogra}},
  \bibinfo{author}{\bibfnamefont{B.}~\bibnamefont{Loer}},
  \bibinfo{author}{\bibfnamefont{F.}~\bibnamefont{Vasconcelos}},
  \bibinfo{author}{\bibfnamefont{D.~K.} \bibnamefont{Kim}},
  \bibinfo{author}{\bibfnamefont{A.~J.} \bibnamefont{Melville}},
  \bibinfo{author}{\bibfnamefont{B.~M.} \bibnamefont{Niedzielski}},
  \bibinfo{author}{\bibfnamefont{J.~L.} \bibnamefont{Yoder}},
  \bibnamefont{et~al.}, \bibinfo{journal}{Nature}
  \textbf{\bibinfo{volume}{584}}, \bibinfo{pages}{551} (\bibinfo{year}{2020}).

\bibitem[{\citenamefont{Wilen et~al.}(2021)\citenamefont{Wilen, Abdullah,
  Kurinsky, Stanford, Cardani, d’Imperio, Tomei, Faoro, Ioffe, Liu
  et~al.}}]{Wilen2021}
\bibinfo{author}{\bibfnamefont{C.}~\bibnamefont{Wilen}},
  \bibinfo{author}{\bibfnamefont{S.}~\bibnamefont{Abdullah}},
  \bibinfo{author}{\bibfnamefont{N.}~\bibnamefont{Kurinsky}},
  \bibinfo{author}{\bibfnamefont{C.}~\bibnamefont{Stanford}},
  \bibinfo{author}{\bibfnamefont{L.}~\bibnamefont{Cardani}},
  \bibinfo{author}{\bibfnamefont{G.}~\bibnamefont{d’Imperio}},
  \bibinfo{author}{\bibfnamefont{C.}~\bibnamefont{Tomei}},
  \bibinfo{author}{\bibfnamefont{L.}~\bibnamefont{Faoro}},
  \bibinfo{author}{\bibfnamefont{L.}~\bibnamefont{Ioffe}},
  \bibinfo{author}{\bibfnamefont{C.}~\bibnamefont{Liu}}, \bibnamefont{et~al.},
  \bibinfo{journal}{Nature} \textbf{\bibinfo{volume}{594}},
  \bibinfo{pages}{369} (\bibinfo{year}{2021}).

\bibitem[{\citenamefont{Cardani et~al.}(2021)\citenamefont{Cardani, Valenti,
  Casali, Catelani, Charpentier, Clemenza, Colantoni, Cruciani, D’Imperio,
  Gironi et~al.}}]{Cardani2021}
\bibinfo{author}{\bibfnamefont{L.}~\bibnamefont{Cardani}},
  \bibinfo{author}{\bibfnamefont{F.}~\bibnamefont{Valenti}},
  \bibinfo{author}{\bibfnamefont{N.}~\bibnamefont{Casali}},
  \bibinfo{author}{\bibfnamefont{G.}~\bibnamefont{Catelani}},
  \bibinfo{author}{\bibfnamefont{T.}~\bibnamefont{Charpentier}},
  \bibinfo{author}{\bibfnamefont{M.}~\bibnamefont{Clemenza}},
  \bibinfo{author}{\bibfnamefont{I.}~\bibnamefont{Colantoni}},
  \bibinfo{author}{\bibfnamefont{A.}~\bibnamefont{Cruciani}},
  \bibinfo{author}{\bibfnamefont{G.}~\bibnamefont{D’Imperio}},
  \bibinfo{author}{\bibfnamefont{L.}~\bibnamefont{Gironi}},
  \bibnamefont{et~al.}, \bibinfo{journal}{Nature Communications}
  \textbf{\bibinfo{volume}{12}}, \bibinfo{pages}{2733} (\bibinfo{year}{2021}).

\bibitem[{\citenamefont{Harrington et~al.}(2024)\citenamefont{Harrington, Li,
  Hays, Van De~Pontseele, Mayer, Pinckney, Contipelli, Gingras, Niedzielski,
  Stickler et~al.}}]{Harrington2024}
\bibinfo{author}{\bibfnamefont{P.~M.} \bibnamefont{Harrington}},
  \bibinfo{author}{\bibfnamefont{M.}~\bibnamefont{Li}},
  \bibinfo{author}{\bibfnamefont{M.}~\bibnamefont{Hays}},
  \bibinfo{author}{\bibfnamefont{W.}~\bibnamefont{Van De~Pontseele}},
  \bibinfo{author}{\bibfnamefont{D.}~\bibnamefont{Mayer}},
  \bibinfo{author}{\bibfnamefont{H.~D.} \bibnamefont{Pinckney}},
  \bibinfo{author}{\bibfnamefont{F.}~\bibnamefont{Contipelli}},
  \bibinfo{author}{\bibfnamefont{M.}~\bibnamefont{Gingras}},
  \bibinfo{author}{\bibfnamefont{B.~M.} \bibnamefont{Niedzielski}},
  \bibinfo{author}{\bibfnamefont{H.}~\bibnamefont{Stickler}},
  \bibnamefont{et~al.}, \bibinfo{journal}{arXiv preprint arXiv:2402.03208}
  (\bibinfo{year}{2024}).

\bibitem[{\citenamefont{Li et~al.}(2024)\citenamefont{Li, Wang, Jiang, Xue,
  Cai, Zhou, Gong, Liu, Zheng, Ma et~al.}}]{li2024}
\bibinfo{author}{\bibfnamefont{X.-G.} \bibnamefont{Li}},
  \bibinfo{author}{\bibfnamefont{J.-H.} \bibnamefont{Wang}},
  \bibinfo{author}{\bibfnamefont{Y.-Y.} \bibnamefont{Jiang}},
  \bibinfo{author}{\bibfnamefont{G.-M.} \bibnamefont{Xue}},
  \bibinfo{author}{\bibfnamefont{X.-X.} \bibnamefont{Cai}},
  \bibinfo{author}{\bibfnamefont{J.}~\bibnamefont{Zhou}},
  \bibinfo{author}{\bibfnamefont{M.}~\bibnamefont{Gong}},
  \bibinfo{author}{\bibfnamefont{Z.-F.} \bibnamefont{Liu}},
  \bibinfo{author}{\bibfnamefont{S.-Y.} \bibnamefont{Zheng}},
  \bibinfo{author}{\bibfnamefont{D.-K.} \bibnamefont{Ma}},
  \bibnamefont{et~al.}, \bibinfo{journal}{arXiv preprint arXiv:2402.04245}
  (\bibinfo{year}{2024}).

\bibitem[{\citenamefont{Thorbeck et~al.}(2023)\citenamefont{Thorbeck, Eddins,
  Lauer, McClure, and Carroll}}]{Thorbeck2023}
\bibinfo{author}{\bibfnamefont{T.}~\bibnamefont{Thorbeck}},
  \bibinfo{author}{\bibfnamefont{A.}~\bibnamefont{Eddins}},
  \bibinfo{author}{\bibfnamefont{I.}~\bibnamefont{Lauer}},
  \bibinfo{author}{\bibfnamefont{D.~T.} \bibnamefont{McClure}},
  \bibnamefont{and} \bibinfo{author}{\bibfnamefont{M.}~\bibnamefont{Carroll}},
  \bibinfo{journal}{PRX Quantum} \textbf{\bibinfo{volume}{4}},
  \bibinfo{pages}{020356} (\bibinfo{year}{2023}).

\bibitem[{\citenamefont{Fowler et~al.}(2012)\citenamefont{Fowler, Mariantoni,
  Martinis, and Cleland}}]{Fowler2012}
\bibinfo{author}{\bibfnamefont{A.~G.} \bibnamefont{Fowler}},
  \bibinfo{author}{\bibfnamefont{M.}~\bibnamefont{Mariantoni}},
  \bibinfo{author}{\bibfnamefont{J.~M.} \bibnamefont{Martinis}},
  \bibnamefont{and} \bibinfo{author}{\bibfnamefont{A.~N.}
  \bibnamefont{Cleland}}, \bibinfo{journal}{Phys. Rev. A}
  \textbf{\bibinfo{volume}{86}}, \bibinfo{pages}{032324}
  (\bibinfo{year}{2012}).

\bibitem[{\citenamefont{Xu et~al.}(2022)\citenamefont{Xu, Seif, Yan, Mannucci,
  Sane, Van~Meter, Cleland, and Jiang}}]{Xu2022}
\bibinfo{author}{\bibfnamefont{Q.}~\bibnamefont{Xu}},
  \bibinfo{author}{\bibfnamefont{A.}~\bibnamefont{Seif}},
  \bibinfo{author}{\bibfnamefont{H.}~\bibnamefont{Yan}},
  \bibinfo{author}{\bibfnamefont{N.}~\bibnamefont{Mannucci}},
  \bibinfo{author}{\bibfnamefont{B.~O.} \bibnamefont{Sane}},
  \bibinfo{author}{\bibfnamefont{R.}~\bibnamefont{Van~Meter}},
  \bibinfo{author}{\bibfnamefont{A.~N.} \bibnamefont{Cleland}},
  \bibnamefont{and} \bibinfo{author}{\bibfnamefont{L.}~\bibnamefont{Jiang}},
  \bibinfo{journal}{Phys. Rev. Lett.} \textbf{\bibinfo{volume}{129}},
  \bibinfo{pages}{240502} (\bibinfo{year}{2022}).

\bibitem[{\citenamefont{Sane et~al.}(2023)\citenamefont{Sane, Meter, and
  Hajdusek}}]{Sane2023}
\bibinfo{author}{\bibfnamefont{B.~O.} \bibnamefont{Sane}},
  \bibinfo{author}{\bibfnamefont{R.~V.} \bibnamefont{Meter}}, \bibnamefont{and}
  \bibinfo{author}{\bibfnamefont{M.}~\bibnamefont{Hajdusek}}, in
  \emph{\bibinfo{booktitle}{2023 IEEE International Conference on Quantum
  Computing and Engineering (QCE)}} (\bibinfo{year}{2023}),
  vol.~\bibinfo{volume}{01}, pp. \bibinfo{pages}{1378--1388}.

\bibitem[{\citenamefont{Suzuki et~al.}(2022)\citenamefont{Suzuki, Sugiyama,
  Arai, Liao, Inoue, and Tanimoto}}]{Suzuki2022}
\bibinfo{author}{\bibfnamefont{Y.}~\bibnamefont{Suzuki}},
  \bibinfo{author}{\bibfnamefont{T.}~\bibnamefont{Sugiyama}},
  \bibinfo{author}{\bibfnamefont{T.}~\bibnamefont{Arai}},
  \bibinfo{author}{\bibfnamefont{W.}~\bibnamefont{Liao}},
  \bibinfo{author}{\bibfnamefont{K.}~\bibnamefont{Inoue}}, \bibnamefont{and}
  \bibinfo{author}{\bibfnamefont{T.}~\bibnamefont{Tanimoto}}, in
  \emph{\bibinfo{booktitle}{2022 55th IEEE/ACM International Symposium on
  Microarchitecture (MICRO)}} (\bibinfo{organization}{IEEE},
  \bibinfo{year}{2022}), pp. \bibinfo{pages}{1110--1125}.

\bibitem[{\citenamefont{Henriques et~al.}(2019)\citenamefont{Henriques,
  Valenti, Charpentier, Lagoin, Gouriou, Mart{\'\i}nez, Cardani, Vignati,
  Gr{\"u}nhaupt, Gusenkova et~al.}}]{Henriques2019}
\bibinfo{author}{\bibfnamefont{F.}~\bibnamefont{Henriques}},
  \bibinfo{author}{\bibfnamefont{F.}~\bibnamefont{Valenti}},
  \bibinfo{author}{\bibfnamefont{T.}~\bibnamefont{Charpentier}},
  \bibinfo{author}{\bibfnamefont{M.}~\bibnamefont{Lagoin}},
  \bibinfo{author}{\bibfnamefont{C.}~\bibnamefont{Gouriou}},
  \bibinfo{author}{\bibfnamefont{M.}~\bibnamefont{Mart{\'\i}nez}},
  \bibinfo{author}{\bibfnamefont{L.}~\bibnamefont{Cardani}},
  \bibinfo{author}{\bibfnamefont{M.}~\bibnamefont{Vignati}},
  \bibinfo{author}{\bibfnamefont{L.}~\bibnamefont{Gr{\"u}nhaupt}},
  \bibinfo{author}{\bibfnamefont{D.}~\bibnamefont{Gusenkova}},
  \bibnamefont{et~al.}, \bibinfo{journal}{Applied Physics Letters}
  \textbf{\bibinfo{volume}{115}}, \bibinfo{pages}{212601}
  (\bibinfo{year}{2019}).

\bibitem[{\citenamefont{Karatsu et~al.}(2019)\citenamefont{Karatsu, Endo,
  Bueno, De~Visser, Barends, Thoen, Murugesan, Tomita, and
  Baselmans}}]{Karatsu2019}
\bibinfo{author}{\bibfnamefont{K.}~\bibnamefont{Karatsu}},
  \bibinfo{author}{\bibfnamefont{A.}~\bibnamefont{Endo}},
  \bibinfo{author}{\bibfnamefont{J.}~\bibnamefont{Bueno}},
  \bibinfo{author}{\bibfnamefont{P.}~\bibnamefont{De~Visser}},
  \bibinfo{author}{\bibfnamefont{R.}~\bibnamefont{Barends}},
  \bibinfo{author}{\bibfnamefont{D.}~\bibnamefont{Thoen}},
  \bibinfo{author}{\bibfnamefont{V.}~\bibnamefont{Murugesan}},
  \bibinfo{author}{\bibfnamefont{N.}~\bibnamefont{Tomita}}, \bibnamefont{and}
  \bibinfo{author}{\bibfnamefont{J.}~\bibnamefont{Baselmans}},
  \bibinfo{journal}{Applied Physics Letters} \textbf{\bibinfo{volume}{114}},
  \bibinfo{pages}{032601} (\bibinfo{year}{2019}).

\bibitem[{\citenamefont{Iaia et~al.}(2022)\citenamefont{Iaia, Ku, Ballard,
  Larson, Yelton, Liu, Patel, McDermott, and Plourde}}]{Iaia2022}
\bibinfo{author}{\bibfnamefont{V.}~\bibnamefont{Iaia}},
  \bibinfo{author}{\bibfnamefont{J.}~\bibnamefont{Ku}},
  \bibinfo{author}{\bibfnamefont{A.}~\bibnamefont{Ballard}},
  \bibinfo{author}{\bibfnamefont{C.}~\bibnamefont{Larson}},
  \bibinfo{author}{\bibfnamefont{E.}~\bibnamefont{Yelton}},
  \bibinfo{author}{\bibfnamefont{C.}~\bibnamefont{Liu}},
  \bibinfo{author}{\bibfnamefont{S.}~\bibnamefont{Patel}},
  \bibinfo{author}{\bibfnamefont{R.}~\bibnamefont{McDermott}},
  \bibnamefont{and} \bibinfo{author}{\bibfnamefont{B.}~\bibnamefont{Plourde}},
  \bibinfo{journal}{Nature Communications} \textbf{\bibinfo{volume}{13}},
  \bibinfo{pages}{6425} (\bibinfo{year}{2022}).

\bibitem[{\citenamefont{Martinis}(2021)}]{Martinis2021}
\bibinfo{author}{\bibfnamefont{J.~M.} \bibnamefont{Martinis}},
  \bibinfo{journal}{npj Quantum Information} \textbf{\bibinfo{volume}{7}},
  \bibinfo{pages}{90} (\bibinfo{year}{2021}).

\bibitem[{\citenamefont{Kelsey et~al.}(2023)\citenamefont{Kelsey, Agnese, Alam,
  Langroudy, Azadbakht, Brandt, Bunker, Cabrera, Chang, Coombes
  et~al.}}]{Kelsey2023}
\bibinfo{author}{\bibfnamefont{M.}~\bibnamefont{Kelsey}},
  \bibinfo{author}{\bibfnamefont{R.}~\bibnamefont{Agnese}},
  \bibinfo{author}{\bibfnamefont{Y.}~\bibnamefont{Alam}},
  \bibinfo{author}{\bibfnamefont{I.~A.} \bibnamefont{Langroudy}},
  \bibinfo{author}{\bibfnamefont{E.}~\bibnamefont{Azadbakht}},
  \bibinfo{author}{\bibfnamefont{D.}~\bibnamefont{Brandt}},
  \bibinfo{author}{\bibfnamefont{R.}~\bibnamefont{Bunker}},
  \bibinfo{author}{\bibfnamefont{B.}~\bibnamefont{Cabrera}},
  \bibinfo{author}{\bibfnamefont{Y.-Y.} \bibnamefont{Chang}},
  \bibinfo{author}{\bibfnamefont{H.}~\bibnamefont{Coombes}},
  \bibnamefont{et~al.}, \bibinfo{journal}{Nuclear Instruments and Methods in
  Physics Research Section A: Accelerators, Spectrometers, Detectors and
  Associated Equipment} \textbf{\bibinfo{volume}{1055}},
  \bibinfo{pages}{168473} (\bibinfo{year}{2023}).

\bibitem[{\citenamefont{Dolan}(1977)}]{Dolan1977}
\bibinfo{author}{\bibfnamefont{G.}~\bibnamefont{Dolan}},
  \bibinfo{journal}{Applied Physics Letters} \textbf{\bibinfo{volume}{31}},
  \bibinfo{pages}{337} (\bibinfo{year}{1977}).

\bibitem[{\citenamefont{Marchegiani et~al.}(2022)\citenamefont{Marchegiani,
  Amico, and Catelani}}]{Marchegiani2022}
\bibinfo{author}{\bibfnamefont{G.}~\bibnamefont{Marchegiani}},
  \bibinfo{author}{\bibfnamefont{L.}~\bibnamefont{Amico}}, \bibnamefont{and}
  \bibinfo{author}{\bibfnamefont{G.}~\bibnamefont{Catelani}},
  \bibinfo{journal}{PRX Quantum} \textbf{\bibinfo{volume}{3}},
  \bibinfo{pages}{040338} (\bibinfo{year}{2022}).

\bibitem[{\citenamefont{Catelani et~al.}(2011)\citenamefont{Catelani,
  Schoelkopf, Devoret, and Glazman}}]{Catelani2011}
\bibinfo{author}{\bibfnamefont{G.}~\bibnamefont{Catelani}},
  \bibinfo{author}{\bibfnamefont{R.~J.} \bibnamefont{Schoelkopf}},
  \bibinfo{author}{\bibfnamefont{M.~H.} \bibnamefont{Devoret}},
  \bibnamefont{and} \bibinfo{author}{\bibfnamefont{L.~I.}
  \bibnamefont{Glazman}}, \bibinfo{journal}{Phys. Rev. B}
  \textbf{\bibinfo{volume}{84}}, \bibinfo{pages}{064517}
  (\bibinfo{year}{2011}).

\bibitem[{\citenamefont{Kaplan et~al.}(1976)\citenamefont{Kaplan, Chi,
  Langenberg, Chang, Jafarey, and Scalapino}}]{Kaplan1976}
\bibinfo{author}{\bibfnamefont{S.~B.} \bibnamefont{Kaplan}},
  \bibinfo{author}{\bibfnamefont{C.}~\bibnamefont{Chi}},
  \bibinfo{author}{\bibfnamefont{D.}~\bibnamefont{Langenberg}},
  \bibinfo{author}{\bibfnamefont{J.-J.} \bibnamefont{Chang}},
  \bibinfo{author}{\bibfnamefont{S.}~\bibnamefont{Jafarey}}, \bibnamefont{and}
  \bibinfo{author}{\bibfnamefont{D.}~\bibnamefont{Scalapino}},
  \bibinfo{journal}{Phys. Rev. B} \textbf{\bibinfo{volume}{14}},
  \bibinfo{pages}{4854} (\bibinfo{year}{1976}).

\bibitem[{\citenamefont{Ullom et~al.}(1998)\citenamefont{Ullom, Fisher, and
  Nahum}}]{Ullom1998}
\bibinfo{author}{\bibfnamefont{J.}~\bibnamefont{Ullom}},
  \bibinfo{author}{\bibfnamefont{P.}~\bibnamefont{Fisher}}, \bibnamefont{and}
  \bibinfo{author}{\bibfnamefont{M.}~\bibnamefont{Nahum}},
  \bibinfo{journal}{Applied Physics Letters} \textbf{\bibinfo{volume}{73}},
  \bibinfo{pages}{2494} (\bibinfo{year}{1998}).

\bibitem[{\citenamefont{Nsanzineza and Plourde}(2014)}]{Nsanzineza2014}
\bibinfo{author}{\bibfnamefont{I.}~\bibnamefont{Nsanzineza}} \bibnamefont{and}
  \bibinfo{author}{\bibfnamefont{B.~L.~T.} \bibnamefont{Plourde}},
  \bibinfo{journal}{Phys. Rev. Lett.} \textbf{\bibinfo{volume}{113}},
  \bibinfo{pages}{117002} (\bibinfo{year}{2014}).

\bibitem[{\citenamefont{Wang et~al.}(2014)\citenamefont{Wang, Gao, Pop, Vool,
  Axline, Brecht, Heeres, Frunzio, Devoret, Catelani et~al.}}]{Wang2014}
\bibinfo{author}{\bibfnamefont{C.}~\bibnamefont{Wang}},
  \bibinfo{author}{\bibfnamefont{Y.~Y.} \bibnamefont{Gao}},
  \bibinfo{author}{\bibfnamefont{I.~M.} \bibnamefont{Pop}},
  \bibinfo{author}{\bibfnamefont{U.}~\bibnamefont{Vool}},
  \bibinfo{author}{\bibfnamefont{C.}~\bibnamefont{Axline}},
  \bibinfo{author}{\bibfnamefont{T.}~\bibnamefont{Brecht}},
  \bibinfo{author}{\bibfnamefont{R.~W.} \bibnamefont{Heeres}},
  \bibinfo{author}{\bibfnamefont{L.}~\bibnamefont{Frunzio}},
  \bibinfo{author}{\bibfnamefont{M.~H.} \bibnamefont{Devoret}},
  \bibinfo{author}{\bibfnamefont{G.}~\bibnamefont{Catelani}},
  \bibnamefont{et~al.}, \bibinfo{journal}{Nature Communications}
  \textbf{\bibinfo{volume}{5}}, \bibinfo{pages}{5836} (\bibinfo{year}{2014}).

\bibitem[{\citenamefont{Diamond et~al.}(2022)\citenamefont{Diamond, Fatemi,
  Hays, Nho, Kurilovich, Connolly, Joshi, Serniak, Frunzio, Glazman
  et~al.}}]{Diamond2022}
\bibinfo{author}{\bibfnamefont{S.}~\bibnamefont{Diamond}},
  \bibinfo{author}{\bibfnamefont{V.}~\bibnamefont{Fatemi}},
  \bibinfo{author}{\bibfnamefont{M.}~\bibnamefont{Hays}},
  \bibinfo{author}{\bibfnamefont{H.}~\bibnamefont{Nho}},
  \bibinfo{author}{\bibfnamefont{P.~D.} \bibnamefont{Kurilovich}},
  \bibinfo{author}{\bibfnamefont{T.}~\bibnamefont{Connolly}},
  \bibinfo{author}{\bibfnamefont{V.~R.} \bibnamefont{Joshi}},
  \bibinfo{author}{\bibfnamefont{K.}~\bibnamefont{Serniak}},
  \bibinfo{author}{\bibfnamefont{L.}~\bibnamefont{Frunzio}},
  \bibinfo{author}{\bibfnamefont{L.~I.} \bibnamefont{Glazman}},
  \bibnamefont{et~al.}, \bibinfo{journal}{PRX Quantum}
  \textbf{\bibinfo{volume}{3}}, \bibinfo{pages}{040304} (\bibinfo{year}{2022}).

\bibitem[{\citenamefont{Leonard et~al.}(2019)\citenamefont{Leonard, Beck,
  Nelson, Christensen, Thorbeck, Howington, Opremcak, Pechenezhskiy, Dodge,
  Dupuis et~al.}}]{Leonard2019}
\bibinfo{author}{\bibfnamefont{E.}~\bibnamefont{Leonard}},
  \bibinfo{author}{\bibfnamefont{M.~A.} \bibnamefont{Beck}},
  \bibinfo{author}{\bibfnamefont{J.}~\bibnamefont{Nelson}},
  \bibinfo{author}{\bibfnamefont{B.~G.} \bibnamefont{Christensen}},
  \bibinfo{author}{\bibfnamefont{T.}~\bibnamefont{Thorbeck}},
  \bibinfo{author}{\bibfnamefont{C.}~\bibnamefont{Howington}},
  \bibinfo{author}{\bibfnamefont{A.}~\bibnamefont{Opremcak}},
  \bibinfo{author}{\bibfnamefont{I.~V.} \bibnamefont{Pechenezhskiy}},
  \bibinfo{author}{\bibfnamefont{K.}~\bibnamefont{Dodge}},
  \bibinfo{author}{\bibfnamefont{N.~P.} \bibnamefont{Dupuis}},
  \bibnamefont{et~al.}, \bibinfo{journal}{Phys. Rev. Appl.}
  \textbf{\bibinfo{volume}{11}}, \bibinfo{pages}{014009}
  (\bibinfo{year}{2019}).

\bibitem[{\citenamefont{Liu et~al.}(2023{\natexlab{a}})\citenamefont{Liu,
  Ballard, Olaya, Schmidt, Biesecker, Lucas, Ullom, Patel, Rafferty, Opremcak
  et~al.}}]{Liu2023}
\bibinfo{author}{\bibfnamefont{C.}~\bibnamefont{Liu}},
  \bibinfo{author}{\bibfnamefont{A.}~\bibnamefont{Ballard}},
  \bibinfo{author}{\bibfnamefont{D.}~\bibnamefont{Olaya}},
  \bibinfo{author}{\bibfnamefont{D.}~\bibnamefont{Schmidt}},
  \bibinfo{author}{\bibfnamefont{J.}~\bibnamefont{Biesecker}},
  \bibinfo{author}{\bibfnamefont{T.}~\bibnamefont{Lucas}},
  \bibinfo{author}{\bibfnamefont{J.}~\bibnamefont{Ullom}},
  \bibinfo{author}{\bibfnamefont{S.}~\bibnamefont{Patel}},
  \bibinfo{author}{\bibfnamefont{O.}~\bibnamefont{Rafferty}},
  \bibinfo{author}{\bibfnamefont{A.}~\bibnamefont{Opremcak}},
  \bibnamefont{et~al.}, \bibinfo{journal}{PRX Quantum}
  \textbf{\bibinfo{volume}{4}}, \bibinfo{pages}{030310}
  (\bibinfo{year}{2023}{\natexlab{a}}).

\bibitem[{\citenamefont{Rist{\`e} et~al.}(2013)\citenamefont{Rist{\`e},
  Bultink, Tiggelman, Schouten, Lehnert, and DiCarlo}}]{Riste2013}
\bibinfo{author}{\bibfnamefont{D.}~\bibnamefont{Rist{\`e}}},
  \bibinfo{author}{\bibfnamefont{C.}~\bibnamefont{Bultink}},
  \bibinfo{author}{\bibfnamefont{M.}~\bibnamefont{Tiggelman}},
  \bibinfo{author}{\bibfnamefont{R.}~\bibnamefont{Schouten}},
  \bibinfo{author}{\bibfnamefont{K.}~\bibnamefont{Lehnert}}, \bibnamefont{and}
  \bibinfo{author}{\bibfnamefont{L.}~\bibnamefont{DiCarlo}},
  \bibinfo{journal}{Nature Communications} \textbf{\bibinfo{volume}{4}},
  \bibinfo{pages}{1913} (\bibinfo{year}{2013}).

\bibitem[{\citenamefont{Christensen et~al.}(2019)\citenamefont{Christensen,
  Wilen, Opremcak, Nelson, Schlenker, Zimonick, Faoro, Ioffe, Rosen, DuBois
  et~al.}}]{Christensen2019}
\bibinfo{author}{\bibfnamefont{B.}~\bibnamefont{Christensen}},
  \bibinfo{author}{\bibfnamefont{C.}~\bibnamefont{Wilen}},
  \bibinfo{author}{\bibfnamefont{A.}~\bibnamefont{Opremcak}},
  \bibinfo{author}{\bibfnamefont{J.}~\bibnamefont{Nelson}},
  \bibinfo{author}{\bibfnamefont{F.}~\bibnamefont{Schlenker}},
  \bibinfo{author}{\bibfnamefont{C.}~\bibnamefont{Zimonick}},
  \bibinfo{author}{\bibfnamefont{L.}~\bibnamefont{Faoro}},
  \bibinfo{author}{\bibfnamefont{L.}~\bibnamefont{Ioffe}},
  \bibinfo{author}{\bibfnamefont{Y.}~\bibnamefont{Rosen}},
  \bibinfo{author}{\bibfnamefont{J.}~\bibnamefont{DuBois}},
  \bibnamefont{et~al.}, \bibinfo{journal}{Phys. Rev. B}
  \textbf{\bibinfo{volume}{100}}, \bibinfo{pages}{140503}
  (\bibinfo{year}{2019}).

\bibitem[{\citenamefont{Kurter et~al.}(2022)\citenamefont{Kurter, Murray,
  Gordon, Wymore, Sandberg, Shelby, Eddins, Adiga, Finck, Rivera
  et~al.}}]{Kurter2021}
\bibinfo{author}{\bibfnamefont{C.}~\bibnamefont{Kurter}},
  \bibinfo{author}{\bibfnamefont{C.~E.} \bibnamefont{Murray}},
  \bibinfo{author}{\bibfnamefont{R.~T.} \bibnamefont{Gordon}},
  \bibinfo{author}{\bibfnamefont{B.~B.} \bibnamefont{Wymore}},
  \bibinfo{author}{\bibfnamefont{M.}~\bibnamefont{Sandberg}},
  \bibinfo{author}{\bibfnamefont{R.~M.} \bibnamefont{Shelby}},
  \bibinfo{author}{\bibfnamefont{A.}~\bibnamefont{Eddins}},
  \bibinfo{author}{\bibfnamefont{V.~P.} \bibnamefont{Adiga}},
  \bibinfo{author}{\bibfnamefont{A.~D.~K.} \bibnamefont{Finck}},
  \bibinfo{author}{\bibfnamefont{E.}~\bibnamefont{Rivera}},
  \bibnamefont{et~al.}, \bibinfo{journal}{npj Quantum Inf}
  \textbf{\bibinfo{volume}{8}}, \bibinfo{pages}{31} (\bibinfo{year}{2022}).

\bibitem[{\citenamefont{Serniak et~al.}(2018)\citenamefont{Serniak, Hays,
  de~Lange, Diamond, Shankar, Burkhart, Frunzio, Houzet, and
  Devoret}}]{Serniak2018}
\bibinfo{author}{\bibfnamefont{K.}~\bibnamefont{Serniak}},
  \bibinfo{author}{\bibfnamefont{M.}~\bibnamefont{Hays}},
  \bibinfo{author}{\bibfnamefont{G.}~\bibnamefont{de~Lange}},
  \bibinfo{author}{\bibfnamefont{S.}~\bibnamefont{Diamond}},
  \bibinfo{author}{\bibfnamefont{S.}~\bibnamefont{Shankar}},
  \bibinfo{author}{\bibfnamefont{L.}~\bibnamefont{Burkhart}},
  \bibinfo{author}{\bibfnamefont{L.}~\bibnamefont{Frunzio}},
  \bibinfo{author}{\bibfnamefont{M.}~\bibnamefont{Houzet}}, \bibnamefont{and}
  \bibinfo{author}{\bibfnamefont{M.}~\bibnamefont{Devoret}},
  \bibinfo{journal}{Phys. Rev. Lett.} \textbf{\bibinfo{volume}{121}},
  \bibinfo{pages}{157701} (\bibinfo{year}{2018}).

\bibitem[{\citenamefont{Pan et~al.}(2022)\citenamefont{Pan, Zhou, Yuan, Nie,
  Wei, Zhang, Li, Liu, Jiang, Catelani et~al.}}]{Pan2022}
\bibinfo{author}{\bibfnamefont{X.}~\bibnamefont{Pan}},
  \bibinfo{author}{\bibfnamefont{Y.}~\bibnamefont{Zhou}},
  \bibinfo{author}{\bibfnamefont{H.}~\bibnamefont{Yuan}},
  \bibinfo{author}{\bibfnamefont{L.}~\bibnamefont{Nie}},
  \bibinfo{author}{\bibfnamefont{W.}~\bibnamefont{Wei}},
  \bibinfo{author}{\bibfnamefont{L.}~\bibnamefont{Zhang}},
  \bibinfo{author}{\bibfnamefont{J.}~\bibnamefont{Li}},
  \bibinfo{author}{\bibfnamefont{S.}~\bibnamefont{Liu}},
  \bibinfo{author}{\bibfnamefont{Z.~H.} \bibnamefont{Jiang}},
  \bibinfo{author}{\bibfnamefont{G.}~\bibnamefont{Catelani}},
  \bibnamefont{et~al.}, \bibinfo{journal}{Nature Communications}
  \textbf{\bibinfo{volume}{13}}, \bibinfo{pages}{7196} (\bibinfo{year}{2022}).

\bibitem[{\citenamefont{Connolly et~al.}(2024)\citenamefont{Connolly,
  Kurilovich, Diamond, Nho, B\o{}ttcher, Glazman, Fatemi, and
  Devoret}}]{Connolly2024}
\bibinfo{author}{\bibfnamefont{T.}~\bibnamefont{Connolly}},
  \bibinfo{author}{\bibfnamefont{P.~D.} \bibnamefont{Kurilovich}},
  \bibinfo{author}{\bibfnamefont{S.}~\bibnamefont{Diamond}},
  \bibinfo{author}{\bibfnamefont{H.}~\bibnamefont{Nho}},
  \bibinfo{author}{\bibfnamefont{C.~G.~L.} \bibnamefont{B\o{}ttcher}},
  \bibinfo{author}{\bibfnamefont{L.~I.} \bibnamefont{Glazman}},
  \bibinfo{author}{\bibfnamefont{V.}~\bibnamefont{Fatemi}}, \bibnamefont{and}
  \bibinfo{author}{\bibfnamefont{M.~H.} \bibnamefont{Devoret}},
  \bibinfo{journal}{Phys. Rev. Lett.} \textbf{\bibinfo{volume}{132}},
  \bibinfo{pages}{217001} (\bibinfo{year}{2024}).

\bibitem[{\citenamefont{Rafferty et~al.}(2021)\citenamefont{Rafferty, Patel,
  Liu, Abdullah, Wilen, Harrison, and McDermott}}]{Rafferty2021}
\bibinfo{author}{\bibfnamefont{O.}~\bibnamefont{Rafferty}},
  \bibinfo{author}{\bibfnamefont{S.}~\bibnamefont{Patel}},
  \bibinfo{author}{\bibfnamefont{C.}~\bibnamefont{Liu}},
  \bibinfo{author}{\bibfnamefont{S.}~\bibnamefont{Abdullah}},
  \bibinfo{author}{\bibfnamefont{C.}~\bibnamefont{Wilen}},
  \bibinfo{author}{\bibfnamefont{D.}~\bibnamefont{Harrison}}, \bibnamefont{and}
  \bibinfo{author}{\bibfnamefont{R.}~\bibnamefont{McDermott}},
  \bibinfo{journal}{arXiv preprint arXiv:2103.06803}  (\bibinfo{year}{2021}).

\bibitem[{\citenamefont{Liu et~al.}(2024)\citenamefont{Liu, Harrison, Patel,
  Wilen, Rafferty, Shearrow, Ballard, Iaia, Ku, Plourde et~al.}}]{liu2022}
\bibinfo{author}{\bibfnamefont{C.~H.} \bibnamefont{Liu}},
  \bibinfo{author}{\bibfnamefont{D.~C.} \bibnamefont{Harrison}},
  \bibinfo{author}{\bibfnamefont{S.}~\bibnamefont{Patel}},
  \bibinfo{author}{\bibfnamefont{C.~D.} \bibnamefont{Wilen}},
  \bibinfo{author}{\bibfnamefont{O.}~\bibnamefont{Rafferty}},
  \bibinfo{author}{\bibfnamefont{A.}~\bibnamefont{Shearrow}},
  \bibinfo{author}{\bibfnamefont{A.}~\bibnamefont{Ballard}},
  \bibinfo{author}{\bibfnamefont{V.}~\bibnamefont{Iaia}},
  \bibinfo{author}{\bibfnamefont{J.}~\bibnamefont{Ku}},
  \bibinfo{author}{\bibfnamefont{B.~L.~T.} \bibnamefont{Plourde}},
  \bibnamefont{et~al.}, \bibinfo{journal}{Phys. Rev. Lett.}
  \textbf{\bibinfo{volume}{132}}, \bibinfo{pages}{017001}
  (\bibinfo{year}{2024}).

\bibitem[{\citenamefont{Houzet et~al.}(2019)\citenamefont{Houzet, Serniak,
  Catelani, Devoret, and Glazman}}]{Houzet2019}
\bibinfo{author}{\bibfnamefont{M.}~\bibnamefont{Houzet}},
  \bibinfo{author}{\bibfnamefont{K.}~\bibnamefont{Serniak}},
  \bibinfo{author}{\bibfnamefont{G.}~\bibnamefont{Catelani}},
  \bibinfo{author}{\bibfnamefont{M.}~\bibnamefont{Devoret}}, \bibnamefont{and}
  \bibinfo{author}{\bibfnamefont{L.}~\bibnamefont{Glazman}},
  \bibinfo{journal}{Phys. Rev. Lett.} \textbf{\bibinfo{volume}{123}},
  \bibinfo{pages}{107704} (\bibinfo{year}{2019}).

\bibitem[{\citenamefont{Anthony-Petersen
  et~al.}(2022)\citenamefont{Anthony-Petersen, Biekert, Bunker, Chang, Chang,
  Chaplinsky, Fascione, Fink, Garcia-Sciveres, Germond et~al.}}]{Anthony2022}
\bibinfo{author}{\bibfnamefont{R.}~\bibnamefont{Anthony-Petersen}},
  \bibinfo{author}{\bibfnamefont{A.}~\bibnamefont{Biekert}},
  \bibinfo{author}{\bibfnamefont{R.}~\bibnamefont{Bunker}},
  \bibinfo{author}{\bibfnamefont{C.~L.} \bibnamefont{Chang}},
  \bibinfo{author}{\bibfnamefont{Y.-Y.} \bibnamefont{Chang}},
  \bibinfo{author}{\bibfnamefont{L.}~\bibnamefont{Chaplinsky}},
  \bibinfo{author}{\bibfnamefont{E.}~\bibnamefont{Fascione}},
  \bibinfo{author}{\bibfnamefont{C.~W.} \bibnamefont{Fink}},
  \bibinfo{author}{\bibfnamefont{M.}~\bibnamefont{Garcia-Sciveres}},
  \bibinfo{author}{\bibfnamefont{R.}~\bibnamefont{Germond}},
  \bibnamefont{et~al.}, \bibinfo{journal}{arXiv preprint arXiv:2208.02790}
  (\bibinfo{year}{2022}).

\bibitem[{\citenamefont{Ramanathan and Kurinsky}(2020)}]{Ramanathan2020}
\bibinfo{author}{\bibfnamefont{K.}~\bibnamefont{Ramanathan}} \bibnamefont{and}
  \bibinfo{author}{\bibfnamefont{N.}~\bibnamefont{Kurinsky}},
  \bibinfo{journal}{Phys. Rev. D} \textbf{\bibinfo{volume}{102}},
  \bibinfo{pages}{063026} (\bibinfo{year}{2020}).

\bibitem[{\citenamefont{Northrop and Wolfe}(1979)}]{Northrop1979}
\bibinfo{author}{\bibfnamefont{G.}~\bibnamefont{Northrop}} \bibnamefont{and}
  \bibinfo{author}{\bibfnamefont{J.}~\bibnamefont{Wolfe}},
  \bibinfo{journal}{Phys. Rev. Lett.} \textbf{\bibinfo{volume}{43}},
  \bibinfo{pages}{1424} (\bibinfo{year}{1979}).

\bibitem[{\citenamefont{Martinez et~al.}(2019)\citenamefont{Martinez, Cardani,
  Casali, Cruciani, Pettinari, and Vignati}}]{Martinez2019}
\bibinfo{author}{\bibfnamefont{M.}~\bibnamefont{Martinez}},
  \bibinfo{author}{\bibfnamefont{L.}~\bibnamefont{Cardani}},
  \bibinfo{author}{\bibfnamefont{N.}~\bibnamefont{Casali}},
  \bibinfo{author}{\bibfnamefont{A.}~\bibnamefont{Cruciani}},
  \bibinfo{author}{\bibfnamefont{G.}~\bibnamefont{Pettinari}},
  \bibnamefont{and} \bibinfo{author}{\bibfnamefont{M.}~\bibnamefont{Vignati}},
  \bibinfo{journal}{Phys. Rev. Applied} \textbf{\bibinfo{volume}{11}},
  \bibinfo{pages}{064025} (\bibinfo{year}{2019}).

\bibitem[{\citenamefont{Glazman and Catelani}(2021)}]{Glazman2021}
\bibinfo{author}{\bibfnamefont{L.~I.} \bibnamefont{Glazman}} \bibnamefont{and}
  \bibinfo{author}{\bibfnamefont{G.}~\bibnamefont{Catelani}},
  \bibinfo{journal}{SciPost Phys. Lect. Notes} \textbf{\bibinfo{volume}{31}}
  (\bibinfo{year}{2021}).

\bibitem[{\citenamefont{Kaplan}(1979)}]{Kaplan1979}
\bibinfo{author}{\bibfnamefont{S.}~\bibnamefont{Kaplan}},
  \bibinfo{journal}{Journal of Low Temperature Physics}
  \textbf{\bibinfo{volume}{37}}, \bibinfo{pages}{343} (\bibinfo{year}{1979}).

\bibitem[{\citenamefont{Wei}(2022)}]{Wei2022}
\bibinfo{author}{\bibfnamefont{P.}~\bibnamefont{Wei}},
  \emph{\bibinfo{title}{Reflection and Transmission of Elastic Waves at
  Interfaces}} (\bibinfo{publisher}{Springer Nature Singapore},
  \bibinfo{address}{Singapore}, \bibinfo{year}{2022}), pp.
  \bibinfo{pages}{63--150}, ISBN \bibinfo{isbn}{978-981-19-5662-1}.

\bibitem[{\citenamefont{Kittel}(2005)}]{Kittel2005}
\bibinfo{author}{\bibfnamefont{C.}~\bibnamefont{Kittel}},
  \emph{\bibinfo{title}{Introduction to Solid State Physics}}
  (\bibinfo{publisher}{Wiley}, \bibinfo{year}{2005}).

\bibitem[{\citenamefont{Simmons and Wang}(1971)}]{SimmonsWang}
\bibinfo{author}{\bibfnamefont{G.}~\bibnamefont{Simmons}} \bibnamefont{and}
  \bibinfo{author}{\bibfnamefont{H.}~\bibnamefont{Wang}},
  \emph{\bibinfo{title}{Single Crystal Elastic Constants and Calculated
  Aggregate Properties: A Handbook}} (\bibinfo{publisher}{MIT Press},
  \bibinfo{year}{1971}), \bibinfo{edition}{2nd} ed.

\bibitem[{\citenamefont{Carbotte}(1990)}]{Carbotte1990}
\bibinfo{author}{\bibfnamefont{J.~P.} \bibnamefont{Carbotte}},
  \bibinfo{journal}{Rev. Mod. Phys.} \textbf{\bibinfo{volume}{62}},
  \bibinfo{pages}{1027} (\bibinfo{year}{1990}).

\bibitem[{\citenamefont{McMillan and Rowell}(1969)}]{McMillan1969Parks}
\bibinfo{author}{\bibfnamefont{W.~L.} \bibnamefont{McMillan}} \bibnamefont{and}
  \bibinfo{author}{\bibfnamefont{J.~M.} \bibnamefont{Rowell}}, in
  \emph{\bibinfo{booktitle}{Superconductivity}}, edited by
  \bibinfo{editor}{\bibfnamefont{R.~D.} \bibnamefont{Parks}}
  (\bibinfo{publisher}{Dekker}, \bibinfo{address}{New York},
  \bibinfo{year}{1969}), vol.~\bibinfo{volume}{1}.

\bibitem[{\citenamefont{Stassis et~al.}(1979)\citenamefont{Stassis, Arch,
  Harmon, and Wakabayashi}}]{Stassis1979}
\bibinfo{author}{\bibfnamefont{C.}~\bibnamefont{Stassis}},
  \bibinfo{author}{\bibfnamefont{D.}~\bibnamefont{Arch}},
  \bibinfo{author}{\bibfnamefont{B.~N.} \bibnamefont{Harmon}},
  \bibnamefont{and}
  \bibinfo{author}{\bibfnamefont{N.}~\bibnamefont{Wakabayashi}},
  \bibinfo{journal}{Phys. Rev. B} \textbf{\bibinfo{volume}{19}},
  \bibinfo{pages}{181} (\bibinfo{year}{1979}).

\bibitem[{\citenamefont{McMillan}(1968)}]{McMillan1968}
\bibinfo{author}{\bibfnamefont{W.~L.} \bibnamefont{McMillan}},
  \bibinfo{journal}{Phys. Rev.} \textbf{\bibinfo{volume}{167}},
  \bibinfo{pages}{331} (\bibinfo{year}{1968}).

\bibitem[{\citenamefont{Gladstone et~al.}(1969)\citenamefont{Gladstone, Jensen,
  and Schrieffer}}]{Gladstone1969Parks}
\bibinfo{author}{\bibfnamefont{G.}~\bibnamefont{Gladstone}},
  \bibinfo{author}{\bibfnamefont{M.~A.} \bibnamefont{Jensen}},
  \bibnamefont{and} \bibinfo{author}{\bibfnamefont{J.~R.}
  \bibnamefont{Schrieffer}}, in \emph{\bibinfo{booktitle}{Superconductivity}},
  edited by \bibinfo{editor}{\bibfnamefont{R.~D.} \bibnamefont{Parks}}
  (\bibinfo{publisher}{Dekker}, \bibinfo{address}{New York},
  \bibinfo{year}{1969}), vol.~\bibinfo{volume}{2}.

\bibitem[{\citenamefont{Sergeev and Mitin}(2000)}]{Sergeev_2000}
\bibinfo{author}{\bibfnamefont{A.}~\bibnamefont{Sergeev}} \bibnamefont{and}
  \bibinfo{author}{\bibfnamefont{V.}~\bibnamefont{Mitin}},
  \bibinfo{journal}{Europhysics Letters} \textbf{\bibinfo{volume}{51}},
  \bibinfo{pages}{641} (\bibinfo{year}{2000}).

\bibitem[{\citenamefont{Ashcroft and Mermin}(1976)}]{AshcroftMermin}
\bibinfo{author}{\bibfnamefont{N.}~\bibnamefont{Ashcroft}} \bibnamefont{and}
  \bibinfo{author}{\bibfnamefont{N.}~\bibnamefont{Mermin}},
  \emph{\bibinfo{title}{Solid State Physics}} (\bibinfo{publisher}{Saunders
  College Publishing}, \bibinfo{year}{1976}).

\bibitem[{\citenamefont{Liu et~al.}(2023{\natexlab{b}})\citenamefont{Liu, He,
  Niu, Xue, Jiang, Ying, Peng, Maezawa, Lin, Xie et~al.}}]{KLiu2023}
\bibinfo{author}{\bibfnamefont{K.}~\bibnamefont{Liu}},
  \bibinfo{author}{\bibfnamefont{X.}~\bibnamefont{He}},
  \bibinfo{author}{\bibfnamefont{Z.}~\bibnamefont{Niu}},
  \bibinfo{author}{\bibfnamefont{H.}~\bibnamefont{Xue}},
  \bibinfo{author}{\bibfnamefont{W.}~\bibnamefont{Jiang}},
  \bibinfo{author}{\bibfnamefont{L.}~\bibnamefont{Ying}},
  \bibinfo{author}{\bibfnamefont{W.}~\bibnamefont{Peng}},
  \bibinfo{author}{\bibfnamefont{M.}~\bibnamefont{Maezawa}},
  \bibinfo{author}{\bibfnamefont{Z.}~\bibnamefont{Lin}},
  \bibinfo{author}{\bibfnamefont{X.}~\bibnamefont{Xie}}, \bibnamefont{et~al.},
  \bibinfo{journal}{Phys. Rev. B} \textbf{\bibinfo{volume}{108}},
  \bibinfo{pages}{064512} (\bibinfo{year}{2023}{\natexlab{b}}).

\bibitem[{\citenamefont{Bargerbos et~al.}(2023)\citenamefont{Bargerbos,
  Splitthoff, Pita-Vidal, Wesdorp, Liu, Krogstrup, Kouwenhoven, Andersen, and
  Gr\"unhaupt}}]{Bargerbos2023}
\bibinfo{author}{\bibfnamefont{A.}~\bibnamefont{Bargerbos}},
  \bibinfo{author}{\bibfnamefont{L.~J.} \bibnamefont{Splitthoff}},
  \bibinfo{author}{\bibfnamefont{M.}~\bibnamefont{Pita-Vidal}},
  \bibinfo{author}{\bibfnamefont{J.~J.} \bibnamefont{Wesdorp}},
  \bibinfo{author}{\bibfnamefont{Y.}~\bibnamefont{Liu}},
  \bibinfo{author}{\bibfnamefont{P.}~\bibnamefont{Krogstrup}},
  \bibinfo{author}{\bibfnamefont{L.~P.} \bibnamefont{Kouwenhoven}},
  \bibinfo{author}{\bibfnamefont{C.~K.} \bibnamefont{Andersen}},
  \bibnamefont{and}
  \bibinfo{author}{\bibfnamefont{L.}~\bibnamefont{Gr\"unhaupt}},
  \bibinfo{journal}{Phys. Rev. Appl.} \textbf{\bibinfo{volume}{19}},
  \bibinfo{pages}{024014} (\bibinfo{year}{2023}).

\bibitem[{\citenamefont{Friedrich et~al.}(1997)\citenamefont{Friedrich, Segall,
  Gaidis, Wilson, Prober, Szymkowiak, and Moseley}}]{Friedrich1997}
\bibinfo{author}{\bibfnamefont{S.}~\bibnamefont{Friedrich}},
  \bibinfo{author}{\bibfnamefont{K.}~\bibnamefont{Segall}},
  \bibinfo{author}{\bibfnamefont{M.~C.} \bibnamefont{Gaidis}},
  \bibinfo{author}{\bibfnamefont{C.~M.} \bibnamefont{Wilson}},
  \bibinfo{author}{\bibfnamefont{D.~E.} \bibnamefont{Prober}},
  \bibinfo{author}{\bibfnamefont{A.~E.} \bibnamefont{Szymkowiak}},
  \bibnamefont{and} \bibinfo{author}{\bibfnamefont{S.~H.}
  \bibnamefont{Moseley}}, \bibinfo{journal}{Applied Physics Letters}
  \textbf{\bibinfo{volume}{71}}, \bibinfo{pages}{3901} (\bibinfo{year}{1997}).

\bibitem[{\citenamefont{Nevala et~al.}(2012)\citenamefont{Nevala, Chaudhuri,
  Halkosaari, Karvonen, and Maasilta}}]{Nevala2012}
\bibinfo{author}{\bibfnamefont{M.~R.} \bibnamefont{Nevala}},
  \bibinfo{author}{\bibfnamefont{S.}~\bibnamefont{Chaudhuri}},
  \bibinfo{author}{\bibfnamefont{J.}~\bibnamefont{Halkosaari}},
  \bibinfo{author}{\bibfnamefont{J.~T.} \bibnamefont{Karvonen}},
  \bibnamefont{and} \bibinfo{author}{\bibfnamefont{I.~J.}
  \bibnamefont{Maasilta}}, \bibinfo{journal}{Applied Physics Letters}
  \textbf{\bibinfo{volume}{101}}, \bibinfo{pages}{112601}
  (\bibinfo{year}{2012}).

\bibitem[{\citenamefont{Julin and Maasilta}(2016)}]{Julin2016}
\bibinfo{author}{\bibfnamefont{J.~K.} \bibnamefont{Julin}} \bibnamefont{and}
  \bibinfo{author}{\bibfnamefont{I.~J.} \bibnamefont{Maasilta}},
  \bibinfo{journal}{Supercond. Sci. Technol.} \textbf{\bibinfo{volume}{29}},
  \bibinfo{pages}{105003} (\bibinfo{year}{2016}).

\end{thebibliography}

\end{document}